\documentclass[pdflatex,sn-basic]{sn-jnl}

\usepackage{graphicx}%
\usepackage{multirow}%
\usepackage{amsmath,amssymb,amsfonts}%
\usepackage{amsthm}%
\usepackage{mathrsfs}%
\usepackage[title]{appendix}%
\usepackage{xcolor}%
\usepackage{textcomp}%
\usepackage{manyfoot}%
\usepackage{booktabs}%
\usepackage{algorithm}%
\usepackage{algorithmicx}%
\usepackage{algpseudocode}%
\usepackage{listings}%
\usepackage{bm}
\usepackage{amssymb,amsbsy,amsmath,amsfonts,amssymb,amscd}
\usepackage{cancel}
\usepackage{mathtools}
\usepackage{hyperref,empheq}
\theoremstyle{thmstyleone}%
\theoremstyle{thmstyletwo}%

\theoremstyle{thmstylethree}%

\def\ie{{{i.e.} }}  
\def\eg{{{e.g.} }}  

\begin{document}


\title[Article Title]{Phenotype structuring in collective cell migration: a tutorial of mathematical models and methods}


\author[1]{\fnm{Tommaso} \sur{Lorenzi}}\email{tommaso.lorenzi@polito.it}

\author[2]{\fnm{Kevin J.} \sur{Painter}}\email{kevin.painter@polito.it}

\author[3,4]{\fnm{Chiara} \sur{Villa}}\email{chiara.villa@inria.fr}

\affil[1]{\orgdiv{Department of Mathematical Sciences ``G. L. Lagrange''}, \orgname{Politecnico di Torino}, \orgaddress{\street{Corso Duca degli Abruzzi, 24}, \city{Torino}, \postcode{10129}, \country{Italy}}}

\affil[2]{\orgdiv{Dipartimento Interateneo di Scienze, Progetto e Politiche del Territorio}, \orgname{Politecnico di Torino}, \orgaddress{\street{Viale Pier Andrea Mattioli, 39}, \city{Torino}, \postcode{10125}, \country{Italy}}}

\affil[3]{Sorbonne Université, CNRS, Université de Paris, Inria, Laboratoire Jacques-Louis Lions UMR 7598, 75005 Paris, France}

\affil[4]{Université Paris-Saclay, Inria, Centre Inria de Saclay, 91120, Palaiseau, France}


\abstract{Populations are heterogeneous, deviating in numerous ways. Phenotypic diversity refers to the range of traits or characteristics across a population, where for cells this could be the levels of signalling, movement and growth activity, etc. Clearly, the  phenotypic distribution -- and how this changes over time and space -- could be a major determinant of population-level dynamics. For instance, across a cancerous population, variations in movement, growth, and ability to evade death may determine its growth trajectory and response to therapy. In this review, we discuss how classical partial differential equation (PDE) approaches for modelling cellular systems and collective cell migration can be extended to include phenotypic structuring. The resulting non-local models -- which we refer to as phenotype-structured partial differential equations (PS-PDEs) -- form a sophisticated class of models with rich dynamics. We set the scene through a brief history of structured population modelling, and then review the extension of several classic movement models -- including the Fisher-KPP and Keller-Segel equations -- into a PS-PDE form. We proceed with a tutorial-style section on derivation, analysis, and simulation techniques. First, we show a method to formally derive these models from underlying agent-based models. Second, we recount travelling waves in PDE models of spatial spread dynamics and concentration phenomena in non-local PDE models of evolutionary dynamics, and combine the two to deduce phenotypic structuring across travelling waves in PS-PDE models. Third, we discuss numerical methods to simulate PS-PDEs, illustrating with a simple scheme based on the method of lines and noting the finer points of consideration. We conclude with a discussion of future modelling and mathematical challenges.}

\keywords{Phenotype-structured populations, Collective cell dynamics, Cell movement, Non-local PDEs, Travelling waves, Concentration phenomena}

\pacs[MSC Classification]{35C07, 35R09, 92B05, 92C17, 92D25}

\maketitle



\section{Introduction}\label{sec1}

`Although everybody seems to have a good intuitive idea of what is meant by a population, one has to sharpen this concept considerably if one attempts to formulate statements about populations in mathematical language.' The above quote, of Heinz von Foerster \citep{vonfoerster1959}, conveys a reality commonly swept under the carpet within simple definitions for a `population' in some model. Often, we airily define a population -- whether of cells or animals -- but tacitly exclude the variation within that arises due to different ages, sizes, chemistry, genetics, phenotypes, etc. Simplifying a population into a group of more or less identical individuals is convenient.

Suppose we wish to model the evolving distribution of some population -- which we will take to be of cells -- where the individuals can proliferate, die, and move: these could be bacteria within a petri-dish, embryonic cells organising to form a tissue, immune cells during an inflammation response, and so forth. A natural and classical approach is to assume the individuals are (more or less) identical and postulate a reaction-advection-diffusion equation for the density $\rho(t,{\bm x})$, at time $t\geq0$ and spatial position ${\bm x}\in \mathcal{X} \subseteq \mathbb{R}^d$, where $d \geq 1$ is the dimensionality of the physical space,
\begin{equation}\label{dar}
\partial_t \rho + \nabla_{{\bm x}} \cdot [ \, {\bm A_{\bm x}}({t},{\bm x})\,  \rho  - D_{\bm x}({t},{\bm x})\, \nabla_{{\bm x}}\rho ]  = \;F(t,{\bm x})\,.
\end{equation}
Here the dependency on $(t,{\bm x})$ of ${\bm A_{\bm x}}, D_{\bm x}$, and $F$ may also be mediated by $\rho$ itself, introducing non-linearities in the equation. 
The kinetics function $F$ describes proliferation and death, while cell migration is split into two components: an advective term with advective velocity $A_{\bm x}$, to describe directed movements, e.g. from guidance cues in the environment;  a diffusive term with diffusion coefficient $D_{\bm x}$, to describe undirected (random) movements. 
The above can of course be extended to systems of coupled equations to define key additional dependent variables. 
A vast range of models have been built with equations that conform to this structure, including many classical systems found within standard textbooks, e.g. \cite{murray2003mathematical}.   

\begin{figure}[t!]
    \centering
    \includegraphics[width=\textwidth]{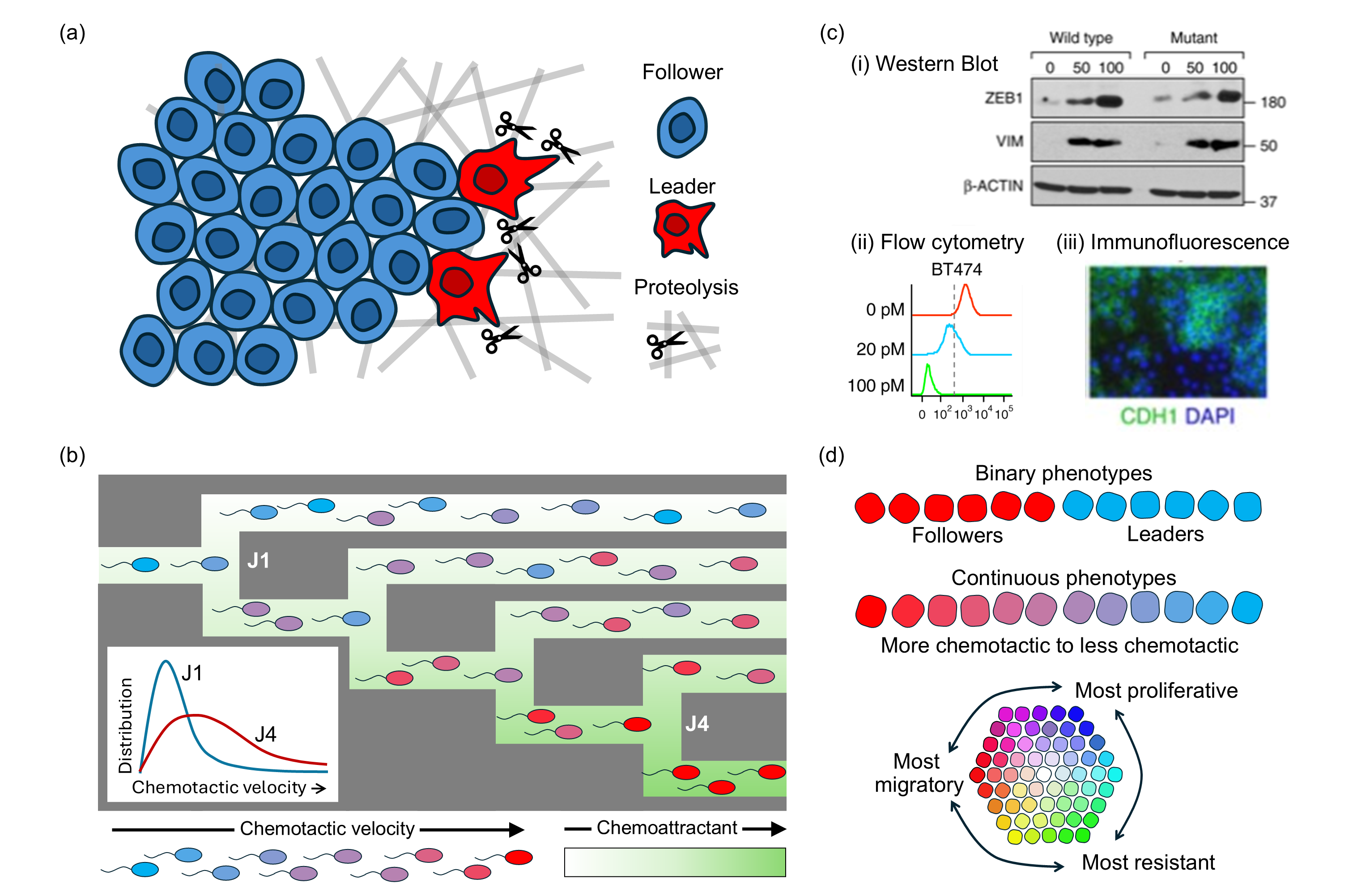} 
    \caption{{\bf Examples of phenotype heterogeneity and its experimental exposition.} (a) Heterogeneity has been intensively studied for its role during cancer invasion, one convenient concept being  a division into `followers' and `leaders'. Leader cells have proteolytic and migratory capabilities that allow them to modify and infiltrate the surrounding extracellular matrix; follower cells expand into the space behind the migrating leaders. For further information, e.g. see \cite{vilchez2021decoding}. (b) An example of how experiments can reveal phenotypic structuring in a population. A T-maze sorts {\em E. coli} cells according to their chemotactic sensitivity: those with stronger responses penetrate deeper into the maze. Figure adapted from the study of \cite{salek2019bacterial}. (c) A range of proteomic methods can provide quantitative data on phenotypic states, including: (i) population-wide average levels of protein expression (e.g. western blot); (ii) protein expression distributions across the population, also from bulk measurements (e.g. flow cytometry); (iii) spatial distributions of protein expression levels, at single-cell and tissue levels (e.g. immunofluorescence). Adapted\protect\footnotemark ~from~\cite{celia2018hysteresis} and licensed under CC BY 4.0 (https://creativecommons.org/licenses/by/4.0/), see the original manuscript for precise information on the depicted data.
    (d) Potential ways of classifying phenotypic variation: (Top) Simple binary variation, where a population is divided into two principal types (e.g. the `follower' or `leader' types in (a)); (Middle) Phenotypes spanning a 1D continuum (e.g. chemotactic sensitivities revealed in T-maze experiments); (Bottom) Phenotypes spanning a higher dimension continuum, e.g. migratory capacity, proliferative capacity, and resistance (i.e. ability to evade death).}
    \label{schematicfigure}
\end{figure}

However, it has long been recognised that populations are rarely (if ever) homogeneous -- some members may move faster, some may proliferate more quickly, etc -- with potentially important consequences. Further, our technological capacity to finely discriminate the various forms of cellular heterogeneity is ever increasing: a repertoire of markers that highlight progression through the cell cycle, allowing an `age distribution' to be tracked \citep{eastman2020palette}; super resolution microscopy \citep{jacquemet2020cell}, that allows cell shape, size, and structure to be recorded with unprecedented detail; protein and mRNA staining techniques (flow cytometry, immunocytochemistry, mass spectronomy, and so forth, see Figure \ref{schematicfigure}(c)) provide information on the distribution of traits at a molecular level \citep{specht2017critical,hu2018single}. In numerous systems, the sharpened picture afforded by these techniques has heightened our understanding into how population distributions are shaped by heterogeneity. To provide a couple of apposite examples: much recent attention has focused on the form of invading cancers, and their structuring from more migratory `leader’ cells at the front to more proliferative `follower’ cells at the rear \citep{vilchez2021decoding}, see Figure \ref{schematicfigure}(a); microfabricated mazes reveal the phenotypic structure within {\em E. coli} bacteria populations, via sorting them according to their ability to respond to chemical gradients by chemotaxis \citep{salek2019bacterial}, see Figure \ref{schematicfigure}(b).  

    \footnotetext{Figures adapted from~\cite{celia2018hysteresis}: (i) Western blot extracted from Figure~2d; (ii) Flow cytometry from Figure~1g; (iii) Immunofluorescence from Figure~1f.}
    
The onus, therefore, is placed on theoreticians to further develop the mathematical theory that can accommodate this more refined level of information: following the above advice of von Foerster, when formulating a model we need precise definitions that fully capture a population’s most critical variables. Adopting a standard nomenclature, we will refer to a population’s distribution across some variable – whether age, size, shape, etc – as a {\em structuring}. Frameworks that permit structure to be incorporated into models date back to the early days of modern mathematical biology. Initially these methods were formulated in order to understand population demographics, leading to age-structured models, but subsequently expanded to account for size structuring and many other variables as deemed relevant. 

We will review briefly this literature, but then turn our attention to mathematical methods and tools that can be used to model and analyse the evolving spatiotemporal distribution of a (cell) population, where the population members also span a (potentially broad) spectrum of {\em phenotypes} or {\em traits}: distinct traits could refer to differences in proliferative potential, capacity to migrate, ability to evade death, and so on. Phenotype variation within such models may range from a (relatively) simplistic binary description, to phenotypes extending across some high-dimensional continuum of states, see Figure \ref{schematicfigure}(c). Models of this nature have become increasingly influential in recent years, a natural response to the large number of experimental studies that have revealed the role of different phenotypes when driving large scale population dynamics.

Retaining the notation $\rho(t,{\bm x})$ to describe the density at time $t$ and position ${\bm x}$, the models we focus on assume a further structuring of the population according to the phenotypic state of the  population members, denoted ${\bm y} \in {\cal {Y}} \subseteq \mathbb{R}^p$, where $p \geq 1$ is the dimensionality of the phenotypic space. We then define $n(t,{\bm x},{\bm y})$ as the {\em phenotype density} at time $t$, position ${\bm x}$, and phenotypic state ${\bm y}$, and note that the (total) {\em density} is then
\[
\rho(t,{\bm x}) := \int_{{\cal {Y}}} n(t,{\bm x},{\bm y}) \, {\rm d}{\bm y}\,.
\]
The models we review here represent the extension of the basic reaction-advection-diffusion framework (\ref{dar}) to accommodate any impacts from phenotypic structuring. First, this involves including dependencies on the phenotypic state ${\bm y}$ in the various terms introduced above, so that the characteristic of a particular phenotype (\eg higher proliferation, lower migration) can be incorporated. Second, it requires the incorporation of additional terms to describe potential transitions through phenotypic space. Specifically, we will review models built on the following structure:
\begin{equation}\label{general}
\begin{cases}
\displaystyle{\partial_t n + \nabla_{{\bm x}} \cdot [ \, {\bm A_{\bm x}}({t},{\bm x},{\bm y})\,  n  - D_{\bm x}({t},{\bm x},{\bm y})\, \nabla_{{\bm x}}n ]}\; \\
\phantom{\displaystyle{\partial_t n }}\displaystyle{\; +  \nabla_{{\bm y}} \cdot [\, {\bm A_{\bm y}}({t},{\bm x},{\bm y})\,  n  - D_{\bm y}({t},{\bm x},{\bm y})\, \nabla_{{\bm y}}n\;] \;   = \;F(t,{\bm x},{\bm y}) \ ,}  \\[5pt]
\displaystyle{\rho(t,{\bm x}) := \int_{{\cal {Y}}} n(t,{\bm x},{\bm y}) \, {\rm d}{\bm y} \ ,}
\end{cases}
\end{equation}
where the dependency on $(t,{\bm x},{\bm y})$ of ${\bm A_{\bm x}}$, $D_{\bm x}$, ${\bm A_{\bm y}}$, $D_{\bm y}$, and $F$ may be mediated by $\rho$ and $n$, introducing non-linearities and non-localities into the equation. 
Analogous to movements through space, phenotypic transitions can potentially occur in either a directed (e.g. factors in the environment that direct cells to switch from one phenotypic state to another) or undirected (e.g. a spontaneous mutation) fashion. Consequently, we introduce the two additional functions ${\bm A}_{\bm y}$ and $D_{\bm y}$ to describe directed and undirected transitions, respectively. Noting that the above equation extends the partial differential equation (PDE) (\ref{dar}) through both additional phenotype structuring and a potential dependence on the (total) density $\rho$, we refer to this as a phenotype-structured partial differential equation (PS-PDE). 
\newcommand{\psipde}{PS-PDE }
\newcommand{\psipdes}{PS-PDEs }

\subsection*{Reading guide}

This article covers both review and tutorial elements and different components may be of varying interest to different readers. We provide the following signpost.
\begin{itemize}
\item Section \ref{sec:history} offers an overview of structured population models. This serves to place the model~\eqref{general} in a broader and historical context, but readers not interested in this aspect can skip this section.
\item Section \ref{pdetopspide} covers three case studies, aimed to show how well-known PDE models for cell movement can be extended into a PS-PDE form. Specifically, we show this for a Fisher-KPP equation, a pressure-based movement model, and a taxis-based model.
\item Section \ref{techniques} forms a tutorial, showing how standard tools and methods used to derive and analyse PDE models can be extended to their phenotype-structured counterparts. Specifically, we show a derivation of these continuum  models from an on lattice agent-based model, illustrate how classical analyses of concentration phenomena and travelling waves can be extended to phenotype-structured models, and provide a straightforward scheme for solving such models. We note that such a scheme has been made publicly available on a repository. 
\item Section \ref{sec:discussion} provides our perspectives on future challenges and directions for this field, covering elements that include applications, modelling, analysis, numerical simulation, and connecting to data.
\end{itemize}


\section{Structured population modelling} \label{sec:history}
Models that incorporate population structuring have been around for more than a century and a broad range of approaches have been developed. Here we offer a short summary, concentrating on works most directly related to the \psipde models described above (for a wider discussion, we refer to \citealt{hoppenstaedt1975mathematical,perthame2006transport,webb1985theory,metz1986dynamics,charlesworth1994,cushing1998,caswell2001matrix,kot2001elements,auger2008structured}).

\subsection{Structured populations and early modelling}

Population counts date almost 6,000 years ago to early Mesopotamian civilisations and, subsequently, cultures from Egypt to China \citep{halacy1980census}. Surveys served purposes from economic (how much tax can we collect?) to militaristic (how many can fight?), but lacked information on population structure: usually, only the numbers of (able) men were recorded with women, children, or the infirm unstated\footnote{Some records from China dating to the 5th century AD did indicate ages, sexes, and relationships of household members, see \citep{durand1960population}.}. Censuses that include information on population structure form a (comparatively) recent undertaking, sometimes part-triggered by population growth concerns: the first (modern) UK census of 1801 took place in the aftermath of the controversy that surrounded Malthus’ 1798 essay on population growth \citep{malthus1798}. Over time, censuses were expanded to collect\footnote{UK Censuses in the latter half of the 19th century even requested numbers of `lunatics', `imbeciles',  or `idiots' living in a household, leading the Registrar General in 1881 to question the value of this information \citep{ons}: ``It is against human nature to expect a mother to admit her young child to be an idiot, however much she may fear this to be true. To acknowledge the fact is to abandon all hope.''.} more information on population structure. 

Descriptions of cellular populations have also undergone increased refinement, with scientific advances as the primary driver. Earliest observations were limited by microscopic power, but by the early 1900s variations in size could be accurately recorded \citep{jennings1908}. Recent developments --  such as live cell imaging with highly specific fluorescence markers \citep{specht2017critical} and a plethora of `multi-omics' methodologies \citep{hu2018single} -- have unlocked our capacity to analyse the inherent heterogeneity present across cells and tissues.  

Faced with these data, the question of how to describe a population is key to modelling. Early models -- such as exponential growth ideas expounded by the likes of \cite{malthus1798}, or the logistic growth model of \cite{verhulst1838notice} -- describe only total population over time, yet these models remain popular when structured information is missing, considered negligible, or dismissed. Models that include structuring date to the work of Alfred Lotka\footnote{Well-known for the Lotka-Volterra equations, but demographics formed his main research area and later worked for the Metropolitan Life Insurance Company. The Lotka-Volterra model can also be regarded as a structured population model, structured into predators and prey.} on demography in the early 1900s \citep{lotka1907art,lotka1907relation,sharpe1911problem}. Lotka's model took the form of a renewal equation and described a population’s evolving age distribution; analysis was used to determine whether a population would evolve to a stable age distribution.

The compartmental approach of \cite{mckendrick1926}\footnote{Anderson McKendrick, celebrated for his work with William Kermack that laid out a theory of mathematical epidemiology, was a physician of remarkable mathematical intuition: ``McKendrick stands in a class by himself. That a man who spent most of his early life in the Indian Medical Service and who was curator of the College of Physicians at Edinburgh afterwards should have anticipated, by so long, so much of the work done later on stochastic processes in this field, is a most remarkable circumstance \citep{irwin1963place}.'' Compartmentalisation into susceptibles, infected, and recovered leads to SIR-type equations and these classic models were introduced in subsequent work with Kermack~\citep{kermack1927}.} supposed a population in which members transition between compartments, each representing a population state (e.g. age, class, infectivity). As a simple example, suppose $N_{t}^{c}$ denotes the number of individuals in compartment $c$ at time $t$. To write down an equation for $N_{t+\tau}^{c}$, where $\tau$ is some time increment, we set $F_t^{c\rightarrow c’}$ to denote the transition function such that an individual has moved from state $c$ at time $t$ into state $c’$ by time $t+\tau$. If we simplify the state space to a single dimension, position compartments at regular intervals on a one-dimensional lattice of spacing $l$, and restrict transitions to neighbouring compartments, then the equation for $N_{t+\tau}^{c}$ derives from simple balancing:  
\begin{equation}\label{mckendrick}
N_{t+\tau}^{c} = N_{t}^{c} F_t^{c\rightarrow c} + N_{t}^{c-l} F_t^{c-l \rightarrow c} + N_{t}^{c+l} F_t^{c+l \rightarrow c}
\,.
\end{equation}
Naturally, the above can be generalised to higher dimensions, transitions between non-adjacent compartments,  etc. This represents a discrete approach to population structure: the element of heterogeneity is discretised into boxes or intervals that span some state space, with individuals entering or exiting as they change their state. Discrete approaches have proven enormously popular when it comes to modelling population structure, a powerful method being the matrix population methods conceived in the 1940s \citep{bernardelli1941population,leslie1945use}. We do not address such methods here, noting that others have covered these ideas in depth, e.g. \cite{caswell2001matrix}. Rather, we shift our attention to continuous approaches, and in particular to structured PDE models related to (\ref{general}).

\subsection{Age-structured models}

\begin{figure}[t!]
    \centering
    \includegraphics[width=\textwidth]{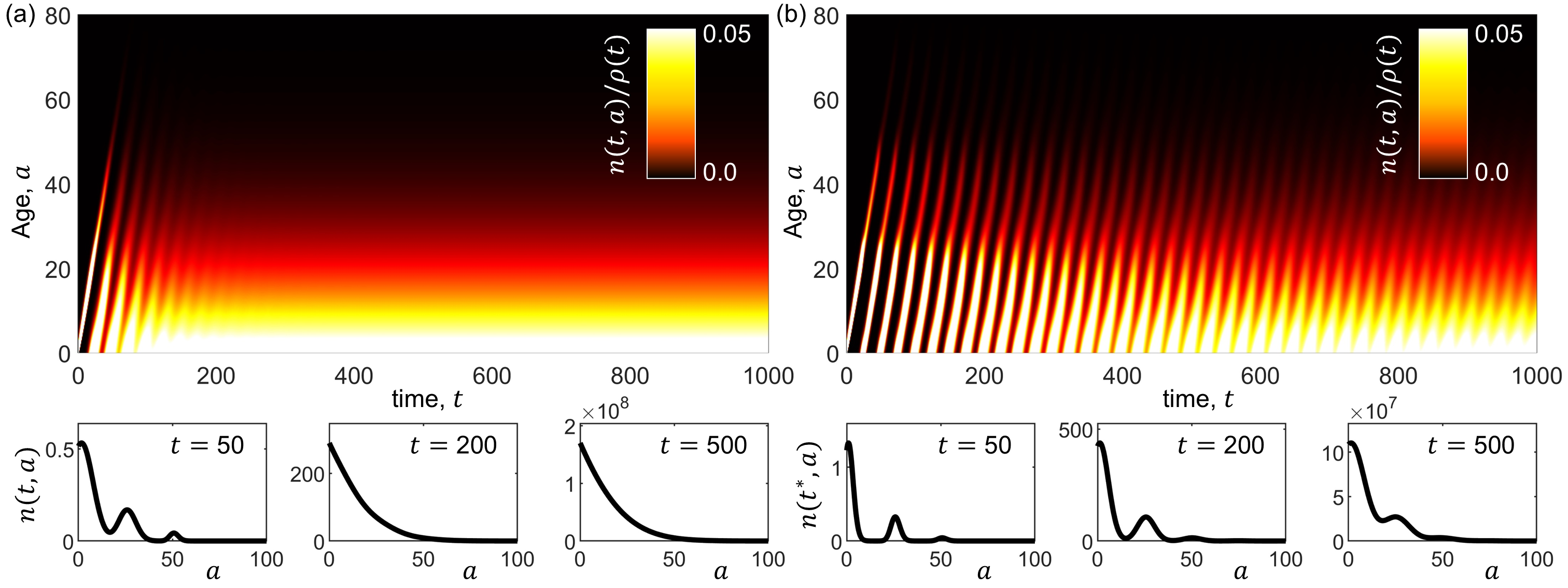}
    \caption{{\bf Illustrative output of the McKendrick-von Foerster equation.} Top panels show the evolving rescaled age distribution, $n(t,a)/\rho(t)$; bottom panels plot the age distribution $n(t,a)$ at select times. Here we assume in (\ref{vonfoerster}) that: (i) the population is initially concentrated towards newborns -- i.e. $n^0(a):=\rho^0 e^{-a}$, where $\rho^0$ is proportional to the initial size of the population (for convenience, we set $\rho_0 = 1$); (ii) the death rate increases linearly with age (i.e. $\mu(t,a,n,\rho) \equiv \mu(a) := 0.001a$); and (iii) births
    result from a Gaussian-like distribution centred about age $a=25$ (i.e. $\beta(t,a,n,\rho) \equiv \beta(a):=(4\exp^{-(a-25)^2/(2\sigma^2)})/(\sqrt{2\pi \sigma^2})$), for (a) dispersed births ($\sigma =4$), and (b) concentrated births ($\sigma =0.25$). In the initial phase, we see a generation that ages with time and new birth ``spikes'' yielding a new generation as the previous generation reaches $a=25$. However, over longer times, the rescaled age distribution reaches a steady-state profile (though this takes longer when births are more concentrated). Note that a steady-state rescaled age distribution does not imply the population size itself is stabilising -- e.g. for these simulations $\rho(t)$ increases exponentially in time.}
    \label{vonfoersterfigure}
\end{figure}

PDE representations of structured populations hold a prominent position within mathematical biology, admitting complex population heterogeneity within a compact framework that can be amenable to analysis. The most well-known model in the field -- the McKendrick-von Foerster equation\footnote{According to the field, sometimes just the McKendrick or von Foerster equation.} – describes the evolving density of a population, $n(t,a)$, structured across age $a \in [0,\infty)$ at time $t \in [0,\infty)$. In a generalised form:
\begin{equation}\label{vonfoerster}
\begin{cases}
	& \partial_t n  +\partial_a n = - \mu(t,a,n,\rho) n\,, \quad a \in (0,\infty) \ ,
  \\[5pt]
	& \displaystyle{\rho(t) := \int_{0}^\infty n(t,a)\, {\rm d}a}\,,
  \\[5pt]
 & \displaystyle{n(0,a) = n^0(a) \ge 0\,},
  \\[5pt]
& \displaystyle{n(t,0) = \int_{0}^\infty \beta(t,a,n,\rho) n(t,a)\, {\rm d}a\,}.
\end{cases}
\end{equation}

In\footnote{In this manuscript we use the notation (\textcolor{blue}{m})$_i$ to refer to the equation on line $i$ for the group with number reference (\textcolor{blue}{m}).}~\eqref{vonfoerster}$_1$, age and time progress in tandem and the right-hand side describes loss or death, with rate $\mu$, where $\rho(t)$ is the (total) population size at time $t$. The condition at $t=0$ is the initial age distribution of the population, i.e. some non-negative function on $[0,\infty)$. The condition at $a=0$ represents births, with $\beta$ the birth rate due to individuals of age $a$. For cell proliferation $\beta$ and $\mu$ may be linked: for example, $\beta = 2 \mu$ if a proliferation leads to two identical daughter cells -- we refer to \cite{perthame2006transport} for more discussion on applications of similar models to cell populations. The model~(\ref{vonfoerster}) is of form (\ref{general}), but with no space dependency and linear advection describing state transitioning. Dynamics predict the evolving age structure; for example, in Figure \ref{vonfoersterfigure} a formulation is shown whereby early waves of distinct generations eventually give way to a stable distribution (but not necessarily stable population size).

The different names given to (\ref{vonfoerster}) highlight its repeated discovery: beyond those of \cite{mckendrick1926} and \cite{vonfoerster1959}\footnote{Heinz von Foerster was a biophysicist with wide-ranging interests, including population growth; his ``Doomsday equation’’ predicted that human population growth rate would become infinite (on November 13th, 2026) \citep{von1960doomsday}. Tongue was firmly in cheek here, as the exact day coincided with what would have been his birthday.} it was derived by \cite{kendall1949} and \cite{scherbaum1957cell}. Different derivations were slightly different in form -- for example, a range from $\mu \equiv 0$ in \citep{scherbaum1957cell} to a non-local dependency $\mu(\rho)$ in~\citep{gurtin1974}. The derivation in \citep{mckendrick1926} is the first documented, where age-structuring was used to illustrate\footnote{Remarkably, the study of \cite{mckendrick1926} used more than ten case applications, often based on data, in examples that included: house to house cancer incidence rates in Luckau, Germany; cholera epidemics in India; bacteria ingestion by leukocytes; infection and relapse in malaria; and, bacteria-antibody conflict, amongst several others.} the compartmental approach when both time and the population state variable could be regarded as continuous. The state space then represents age (we change $c$ to $a$ in (\ref{mckendrick})), age and time are absolutely correlated, and ageing is (unfortunately) a one-way process: we can therefore set $l \equiv \tau$ and $F_t^{a \rightarrow a} =  F_t^{a+\tau \rightarrow a} =0$. Not all individuals progress to the next age class (as some will die), leading to $F_a^{a-\tau \rightarrow a} = 1-\mu \tau $ where $\mu \tau$ is the fraction of the population with age $a-\tau$ at time $t$ that die before they reach age $a$ at time $t+\tau$ . Substituting these into (\ref{mckendrick}), expanding about $(t,a)$, and taking the limit as $\tau\rightarrow 0$, one formally arrives at (\ref{vonfoerster})$_1$ under continuity assumptions.

As noted, a derivation of (\ref{vonfoerster}) is also found in \cite{kendall1949}, for a stochastic birth-death process that accounted for age structuring of a population. These earlier derivations went somewhat under the radar (at least in the context of growth kinetics), with wider awareness not emerging until the 1960s -- primarily to describe cellular growth -- following the derivations of \cite{scherbaum1957cell} and \cite{vonfoerster1959}. The naming after von Foerster appears to originate in \cite{trucco1965a,trucco1965b}, and the 1960s witnessed its wider application to problems in cellular kinetics (e.g. \citealt{fredrickson1963continuous,trucco1965a,trucco1965b,rubinow1968maturity}) and population demographics (e.g. \citealt{hoppenstaedt1975mathematical,keyfitz1972mathematics}). 

\subsection{Age- and space-structured models}
Age is one way a population can be structured, but model (\ref{vonfoerster}) can be straightforwardly extended to include other forms of heterogeneity. One approach would be to add a discrete structuring, such as coupling two equations of type~\eqref{vonfoerster}$_1$ to account for female and male members of a population \citep{fredrickson1971mathematical,keyfitz1972mathematics}. Further continuous structure can be added, with early extensions to size and age structures proposed by \cite{fredrickson1963continuous,bell1967cell}, and \cite{sinko1967new}. Generalisations to a $p$-dimensional ``physiological state vector’’ were first made in \citep{fredrickson1967statistics}, and we refer to \citep{gyllenberg1983stability,webb1985theory,tucker1988nonlinear,webb2008population} for further discussion of high-dimensional structured population models. Other structuring variables may not obey the strict correlation of age with time, for example size may increase or decrease over time; as such, different transport terms may be needed to describe movements through the structuring variable state space.

Physical space is frequently one of the most crucial population structuring variables. Of course, mathematical models that account for spatiotemporal evolution have a long and illustrious history, frequently using the reaction-advection-diffusion form~(\ref{dar}); we rarely refer to these as structured population models, though. Particularly pertinent here, however, are extensions of age-structured models to include spatial movement. These were first introduced by Gurtin and others \citep{gurtin1973,gurtin1977,gurtin1981}, who added a diffusive-type spatial movement to an age-structured population. Starting with model (\ref{vonfoerster}), but extending to a population $n(t,{\bm x},a)$ for time $t$, position ${\bm x}\in\mathcal{X}\subset \mathbb{R}^d$ and age $a\in[0,\infty)$, yields 
\begin{equation*}
\begin{cases}
	& \partial_t n + \partial_a n - \nabla_{\bm x} \cdot \left[D(a)\nabla n \right] = - \mu (t,{\bm x},a,n,\rho) n\,, \quad {\bm x} \in \mathcal{X} \ , \; a \in (0,\infty) \ ,
  \\[5pt]
	&\displaystyle{\rho(t,{\bm x}) := \int_0^{\infty} n(t,{\bm x},a)\, {\rm d}a}\,, 
  \\[5pt]
 & n(0,{\bm x},a) = n^0({\bm x},a)\,,
  \\[5pt]
& \displaystyle{n(t,{\bm x},0) = \int_0^\infty \beta(t,{\bm x},a,n,\rho) n(t,{\bm x},a)\, {\rm d}a}\,, \quad {\bm x} \in \mathcal{X}\,,
  \\[5pt]
  & {\boldsymbol{\nu}} \cdot D(a)\nabla n = q(t,{\bm x},a), \quad {\bm x} \in \partial \mathcal{X} \ , \; a \in (0,\infty) ,
\end{cases}
\end{equation*}
where ${\boldsymbol{\nu}}$ is the outer unit normal on the domain boundary $\partial \mathcal{X}$. The above is of \psipde form, adapted and furnished with boundary and initial conditions relevant for a population structured in age and space. Structured models that include variation in space have subsequently received a huge amount of attention: a rather indiscriminate list includes \citep{garroni1982age,webb1982diffusive,langlais1985nonlinear,fitzgibbon1996diffusive,ainseba2000,al2002monotone,ayati2006structured,ayati2006computational,delgado2006nonlinear,dyson2007age,gandolfi2011age,domschke2017structured,fitzgibbon2018vector,deng2020analysis,kang2021nonlinear}, and involve applications from ecology to cancer growth. Indeed, the primary focus of the rest of this review will be on population models where the structuring is in space and phenotype. In this regard, we also note that variants of (\ref{vonfoerster}) for age- and phenotype-structured populations have recently been considered~\citep{nordmann2018dynamics}. 
 
\subsection{Phenotype-structured models}
The word phenotype refers to the observable features or characteristics of a cell or organism, a term first proposed by \cite{johannsen1911genotype} to distinguish an individual's genetic material (the genotype) from its result (the phenotype). Used for cell populations it can refer to physical characteristics (such as shape, size), molecular (such as gene/mRNA/protein expression), and behaviours (such as migration, proliferation). With respect to the historical models above, variations in age (progression through cell cycle), size, and physiological state could all relate to phenotypic structuring. 

The drivers of phenotypic heterogeneity are diverse. The environment is one important factor -- e.g. as demonstrated by the distinct patterns of metabolic activity that emerge in a microbial population subject to diverse nutrient levels \citep{schreiber2020environmental}. Interactions between individuals can also drive heterogeneity: a `salt-and-pepper’ expression pattern forms across certain embryonic cell populations in development, driven by Delta/Notch-mediated interactions between adjacent cells \citep{shaya2011notch}. Even in the absence of other drivers, phenotypic diversity can arise in clonal cell populations solely through stochastic variation in gene expression \citep{elowitz2002stochastic}. 

Growing appreciation for phenotypic variation has resulted in numerous mathematical models that can be used to understand the impact it has on the spatiotemporal evolution of cellular systems. At a broad level, these attempts can be grouped into two main classes, each with two subclasses: 
\begin{itemize}
	\item Discrete population models, with discrete or continuous phenotypic states;
 \item[]
	\item Continuous populations models, with discrete or continuous phenotypic states.
\end{itemize}
Discrete population approaches are (usually) predicated on agent-based models (ABMs), where each agent is an individual cell with an assigned phenotypic state; this state is then a determinant of the rules that govern the agent’s behaviour. The flexibility of ABMs, coupled to increased computational power, has led to their rapid expansion in biological modelling (e.g. see \citealt{grimm2005,drasdo2018agent,metzcar2019review,van2018off,wang2015simulating} and references therein); it is far beyond the scope of this review to cover these in depth, so we just confine to broad remarks. Within each agent the phenotypic state could be discrete or continuous and fixed or variable. Under discrete states, an agent is restricted to occupying one of a finite set of states, whereas under continuous states the phenotype spans a continuum. For varying states, an agent’s phenotypic state can change over time, e.g. as a result of interactions with other cells or the environment. For example, this could be enacted through equipping each cell with a dynamical system to describe intracellular signalling; various open source toolkits are available for implementing such ABMs -- e.g. see the review by \cite{metzcar2019review}.

Continuous approaches assume that the population can be represented by a continuum, i.e. as a density distribution. As above, these can be further divided into those with discrete or continuous phenotype states. Under the former, the overall dynamics are represented by a system of (typically coupled) evolution equations, each ODE or PDE describing the evolution of the density of individuals with a certain phenotype. Models of this type have become popular in recent years, in particular to describe how cellular heterogeneity can have an impact on cancer progression, invasion, and treatment (e.g. \citealt{gatenby2003glycolytic,fedotov2007migration,gerlee2012impact,pham2012density,stepien2018traveling,strobl2020mix,crossley2024phenotypic}). Relying on just a few phenotype states -- such as a binary description, see  Figure~\ref{schematicfigure}(d) -- these models fall into the general class of low(ish) order ODE/PDE systems and can be approached analytically with standard techniques (linear stability, bifurcation analyses etc). Continuous models in which the phenotype enters as a continuous structuring variable represent a further step up in model complexity, but also benefit from a capacity to capture the complex heterogeneity often present in a population. Many models of this type fall into the \psipde structure~\eqref{general}, and we will be using the rest of this review to describe these in more detail.

\section{PDEs to PS-PDEs for biological movement} \label{pdetopspide}

\bigskip
The PDE (\ref{dar}) underpins many models and has been tailored to describe various types of movement. We use this section to discuss the extension of some classic PDE models of form (\ref{dar}) into the \psipde form (\ref{general}). 
In cell systems, dynamics are regulated by the interactions between cells and other components of the extracellular environment. For this reason, the dependency on $({t},{\bm x})$ of the terms ${\bm A_{\bm x}}$,  ${D_{\bm x}}$, ${\bm A_{\bm y}}$,  ${D_{\bm y}}$, and $F$ in~\eqref{general} may be mediated by macroscopic quantities, in addition to the density $\rho$, modelling other biotic (\eg densities of cells of other populations) and abiotic (\eg concentrations of soluble molecules and densities of insoluble polymers) factors in the microenvironment. These in turn may satisfy their own evolution equations, but for the sake of conciseness we will not discuss them here. The definitions of ${\bm A_{\bm x}}$ and ${D_{\bm x}}$ will depend on the type of movement one seeks to represent, and in the following we will consider some standard forms corresponding to different cell migration models. Note also that while in this section we retain the notation introduced in Section~\ref{sec1}, considering a $p$-dimensional phenotypic structure ($p\geq1$), the majority of the modelling works in the extant literature cited below consider a one-dimensional phenotypic structure ($p=1$). As a further note, the PDEs and PS-PDEs presented in this section -- when posed on bounded domains -- are subject to zero-flux boundary conditions, unless stated otherwise.

\subsection{Diffusion-based movement models} 

\begin{figure}[t!]
    \centering
    \includegraphics[width=\textwidth]{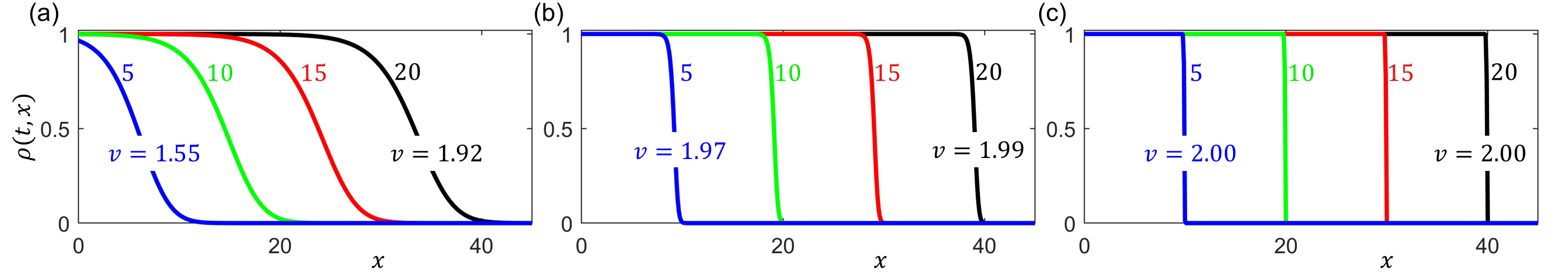}
    \caption{{\bf Illustration of travelling wave like solutions from the Fisher-KPP model.} Illustrative outputs of the Fisher-KPP model~\eqref{FKPP}-\eqref{def:RFKPP}. In particular, evolution towards a travelling wave form is shown for the rescaled PDE~\eqref{FisherKPP:rescaled}, complemented with an initial condition of form~\eqref{FisherKPP:rescaledIC}, with (a) $\varepsilon = 1$, (b) $\varepsilon = 0.1$, (c) $\varepsilon=0.01$. In each case the density $\rho(t,x)$ is plotted at times $t=5$ (blue lines), $t=10$ (green lines), $t=15$ (red lines), and $t=20$ (red lines). We see an evolution of the solutions towards a travelling wave profile with wave speed $c=2$ (numerically calculated wave speeds $v$ reported for each simulation at times $t=5$ and $t=20$). As $\varepsilon \rightarrow 0$, this occurs on a fast timescale and the profile becomes increasingly step-like, from $\rho \equiv 1$ across the wave to $\rho \equiv 0$ ahead of the wave, in agreement with analytical predictions. While travelling waves are usually considered on an unbounded domain, the numerical method requires a bounded domain. For these simulations we have set ${\bm x} \equiv x \in[0,50]$, subject to zero-flux boundary conditions.}\label{FKPPfigure}
\end{figure}

Linear (or Fickian) diffusion describes the effect of Brownian motion at the macroscopic scale (\ie the undirected motion of particles randomly moving in space), and follows from Fick's first law~\citep{fick1855ueber}. The relative simplicity of this form has led to its adoption in many models of biological motion  and is a component of a mathematical biology mainstay: the Fisher-KPP equation, which we adopt as an exemplar. 

Proposed by \cite{fisher1937wave} and \cite{kolmogorov1937study}, the Fisher-KPP equation forms travelling waves for a population that expands about its focal point (see Figure~\ref{FKPPfigure}), and lies at the basis of models for processes that range from ecological invasions to tumour growth. In its textbook statement, linear diffusion is combined with logistic growth so that one defines $D_{\bm x} \equiv D$, ${\bm A_{\bm x}} \equiv 0$, $F:=\rho R(\rho)$ in (\ref{dar}) and obtains
\begin{equation} \label{FKPP}
	\partial_t \rho = D \nabla_{\bm x}^2 \rho + \rho R(\rho)\,, \quad {\bm x} \in \mathcal{X} \ .
\end{equation}
Here the parameter $D \in \mathbb{R}^+$ is the (constant) diffusivity and the function $R$ is the growth rate of the population, which is of the logistic form
\begin{equation} \label{def:RFKPP}
R(\rho) := r \left(1-\dfrac{\rho}{k}\right),
\end{equation}
where $r \in \mathbb{R}^+_*$\footnote{In the remainder of the paper, we will use the notation $\mathbb{R}^+_* := \mathbb{R}^+ \setminus \{0\}$, where $\mathbb{R}^+$ is the set of non-negative real numbers.} is the intrinsic growth rate (i.e. the growth rate when $\rho=0$) and $k \in \mathbb{R}^+_*$ is the local carrying capacity.

A conventional reading of (\ref{FKPP}) suggests a single (i.e. homogeneous) population model, but the original works of \cite{fisher1937wave} and \cite{kolmogorov1937study} were, in fact, for structured (i.e. heterogeneous) populations. For example, Fisher was investigating\footnote{In commenting on this background to (\ref{FKPP}), we should mention the roots to eugenics: \cite{fisher1937wave} appeared in the (discontinued) Annals of Eugenics and Fisher was an editor of the journal. This field became increasingly politicised and notorious across the first decades of the 20th century, so our discussion here is just to highlight that the common interpretation of (\ref{FKPP}) as a single population model had, in its origin, a structured population in mind.} the spread of advantage-gaining mutations. Under this motivation, the function $\rho$ in (\ref{FKPP}) describes the density of the `daughter' population of an allelomorph pair, with the density of the `parent' population set by $k-\rho$ under an assumption that the combined daughter and parent density is constant. The travelling waves that can arise (see Figure \ref{FKPPfigure}) then describe how the daughter population expands to replace the parent. As such, the well-known Fisher-KPP equation -- as originally intended -- described the evolving heterogeneous structure in a population, albeit under a highly simplified scenario. 

\begin{figure}[t!]
    \centering
    \includegraphics[width=\textwidth]{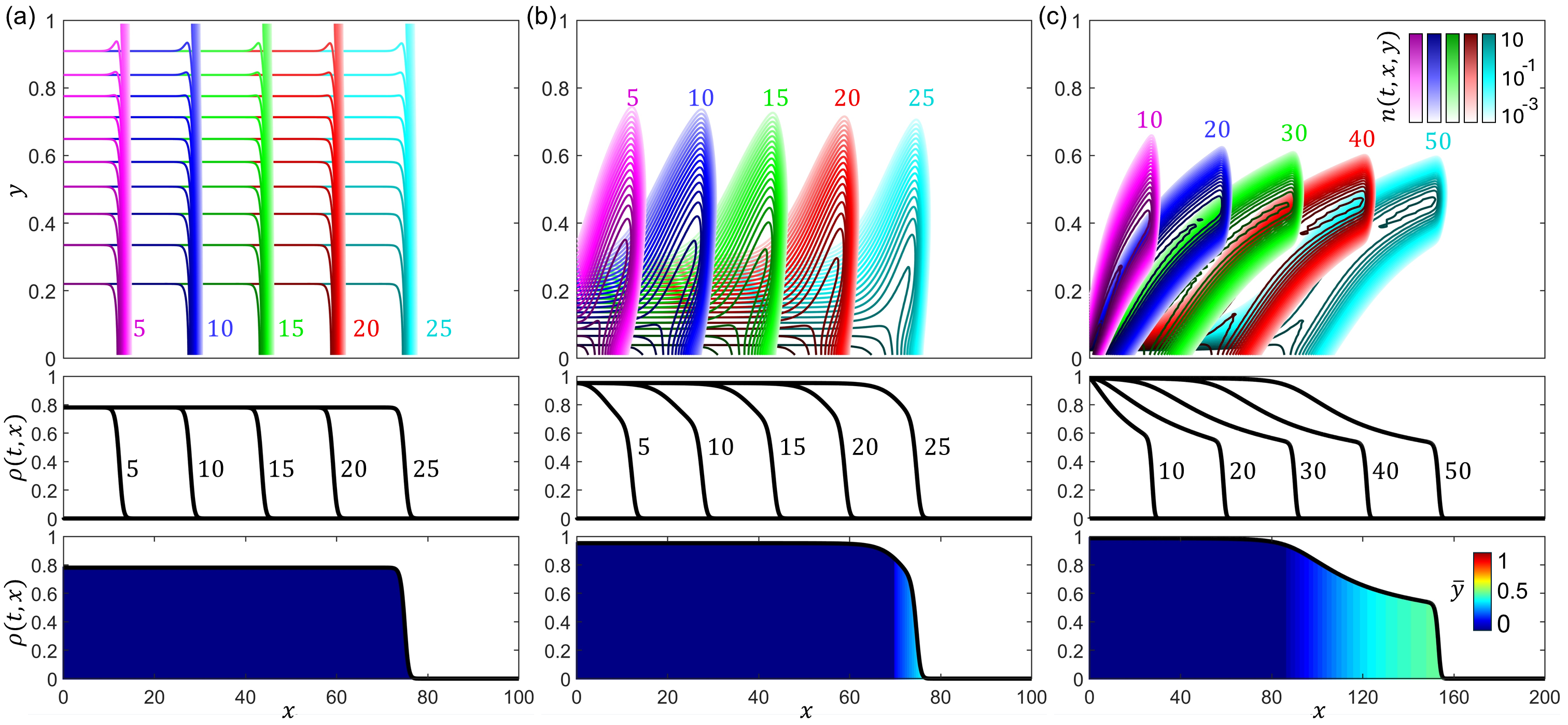}
    \caption{{\bf Illustration of travelling wave like solutions from a \psipde model of diffusion-based movement.} Illustrative output of the PS-PDE model of diffusion-based movement~\eqref{structuredFKPP}. Here, we consider one dimension in each of physical and phenotype space, such that ${\bm x} \equiv x \in[0,L]$ with $L \in \mathbb{R}^+_*$ and ${\bm y} \equiv y \in[0,1]$. Zero-flux conditions are imposed at the boundaries and the population is initially set such that $n(0,x,y)=0.1 e^{-100x}$. We take $D(y):=y$ and define $R(y,\rho) := r(y) - \kappa \rho$, with $r(y) := 10 (1-y)$ and $\kappa=10$. These definitions describe a trade-off in which the most motile population members ($y \approx 1$) have the lowest proliferation rate, while the least motile members ($y \approx 0$) have the highest proliferation rate. Simulations are for (a) $\bar{D}=10^{-1}$, (b) $\bar{D}=10^{-3}$, and (c) $\bar{D}=10^{-5}$. Note that $L=100$ in (a-b) and $L=200$ in (c), due to the longer time it takes to evolve to the travelling wave profile in (c). Top row displays the phenotype density function  $n(t,x,y)$ with the different colour-themes corresponding to different times (colour bars in (c)). Central row displays the density $\rho(t,x)$ at the same times, showing the evolution to travelling wave solutions. Bottom row displays the density $\rho$ at (a-b) $t=25$ and (c) $t=50$, where the colour code indicates the locally prevailing phenotype, $\bar{\bm y}(t,{\bm x}) \equiv \bar{y}(t,x)$, across the wave for (a-b) $t=25$ and (c) $t=50$. Broadly speaking, lowering the diffusivity in phenotype space (i.e. reducing the value of the rate of phenotypic changes $\bar{D}$) leads to an increasingly structured profile, where more mobile members dominate the front of the wave and more proliferative members dominate the rear.}\label{structuredFKPPfigure}
\end{figure}

How can we generalise this to incorporate even greater structuring, i.e. a population continuously structured across some phenotypic space? Let us consider the phenotype density $n(t,{\bm x},{\bm y})$ and allow transitions to take place between the phenotypic states ${\bm {y}} \in {\cal {Y}}$. Taking the PS-PDE form (\ref{general}), it is common to assume that phenotypic changes can be described by a linear diffusion process, so that ${\bm A}_{\bm y} \equiv 0$ and $D_{\bm y} \equiv \bar D$. In cell populations, such a modelling approach is natural, say, if changes are driven by stochastic fluctuations of signalling pathways~\citep{chisholm2016cell}. If the phenotype of population members affects both their motility and proliferation rate, a natural extension of (\ref{FKPP}) is then to the following PS-PDE formulation:
\begin{equation}\label{structuredFKPP}
\begin{cases}
\displaystyle{\partial_t n = D({\bm y}) \nabla_{\bm x}^2 n + \bar{D} \nabla_{\bm y}^2 n + n R({\bm y},\rho) \ , \quad {\bm x} \in \mathcal{X} \ , \; {\bm y} \in \mathcal{Y} \ ,  }\\[5pt]
\displaystyle{\rho(t,{\bm x}) := \int_{{\cal {Y}}} n(t,{\bm x},{\bm y}) \, {\rm d}{\bm y} \ .}
\end{cases}
\end{equation}
In the framework of PS-PDE models of form~\eqref{structuredFKPP}, the parameter $\bar D \in \mathbb{R}^+$ can be regarded as the rate of phenotypic changes. Furthermore, the function $R({\bm {y}},\rho)$ models the net proliferation rate (i.e. the difference between the rate of proliferation and the rate of death) of population members in the phenotypic state ${\bm {y}}$ under the local environmental conditions corresponding to the density $\rho$. Hence, it can be regarded as the fitness function or the fitness landscape of the population~\citep{burger2000mathematical}. 

Depending on the assumptions that are made on the functions and parameters of the model, PS-PDEs of form~\eqref{structuredFKPP} can produce phenotype-structured travelling waves, wherein the locally prevailing phenotype, $\bar{\bm y}$, at different positions -- which we define as the point $\bar{\bm y}(t,{\bm x})$  such that $\displaystyle{n(t,{\bm x},\bar{\bm y}(t,{\bm x})) = \max_{{\bm y} \in \mathcal{Y}} n(t,{\bm x},{\bm y})}$ -- varies across the wave (see Figure~\ref{structuredFKPPfigure}). These can be seen as phenotypically heterogeneous waves of invasion, whereby population members in different phenotypic states dominate different parts of the wave. 

Variants of PS-PDE models of form~\eqref{structuredFKPP} for situations in which the phenotype structure is not linked to the motility of the population members (i.e. when $D({\bm y})\equiv D$) have received attention from the mathematical community for the richness of the properties of their solutions~\citep{arnold2012existence,li2024propagation}. On the other hand, variants of~\eqref{structuredFKPP} for scenarios where the phenotype structure is linked to the motility of the population members but not to the rates at which they proliferate/die, i.e. \psipde models of form 
\begin{equation}\label{eq:CT}
\begin{cases}
\displaystyle{\partial_t n = D({\bm y}) \nabla_{\bm x}^2 n + \bar{D} \nabla_{\bm y}^2 n + n R(\rho)\,,\quad {\bm x} \in \mathcal{X} \ , \; {\bm y} \in \mathcal{Y} \ ,  }\\[5pt]
\displaystyle{\rho(t,{\bm x}) := \int_{{\cal {Y}}} n(t,{\bm x},{\bm y}) \, {\rm d}{\bm y}\,,}
\end{cases}
\end{equation}
have been particularly popular in the study of cane toad invasion in Australia. For instance, considering $\mathcal{Y} \subseteq \mathbb{R}^+$ and thus ${\bm y} \equiv y$, a model of this type with\footnote{Note that the growth rate $R(\rho)$ defined via~\eqref{def:RDCT} is of the logistic form~\eqref{def:RFKPP} with $k=1$. Note also that, under the assumption $\mathcal{Y} \subseteq \mathbb{R}^+$, defining $D$ via~\eqref{def:RDCT} translates into mathematical terms the idea that there is a proportional relationship between the phenotype structuring variable and the motility of the individuals.} 
\begin{equation}
\label{def:RDCT}
D(y):= y \ , \quad R(\rho):=r \left(1-\rho\right)
\end{equation}
was proposed by~\cite{benichou2012front}. More recently, related models that take into account the effect of spatial heterogeneity in the surrounding environment on the proliferation rate~\citep{lam2017integro} and motility~\citep{zamberletti2022spatial} of the population members have also been considered. Over the last decade, the PS-PDE model~\eqref{eq:CT}-\eqref{def:RDCT} has received considerable attention, both from a modelling and an analytical point of view. In particular, from the modelling perspective, it has provided a robust mechanistic explanation for spatial sorting observed in cane toad invasion: highly motile individuals are found to reside at the edge of the invasion front~\citep{benichou2012front,berestycki2015existence,bouin2012invasion,bouin2014travelling,turanova2015model}. From the analytical standpoint, it has become a prototypical example of PS-PDE models that admit accelerating-front solutions
(i.e. travelling-front solutions that accelerate over time)~\citep{berestycki2015existence,bouin2012invasion,bouin2017super}.

Moreover, when spatial dynamics are ignored, i.e. $n(t,{\bm x},{\bm y})\equiv n(t,{\bm y})$, PS-PDE models of form (\ref{structuredFKPP}) reduce to the following well-investigated class of non-local PDE models of evolutionary dynamis~\footnote{It is common to refer to non-local PDE models of form~\eqref{nonlocalFKPP} as mutation-selection models,  e.g. see~\citep{lorenzi2020asymptotic}, or replicator-mutator equations, e.g. see~\citep{alfaro2019evolutionary}, or Lotka-Volterra parabolic equations, e.g. see~\citep{perthame2008dirac}. On the other hand, when $R({\bm {y}},\rho) \equiv R(\rho)$ with $R(\rho)$ of form~\eqref{def:RFKPP} or of a related form, non-local PDE models of this type are usually referred to as non-local Fisher-KPP equation, e.g. see~\citep{berestycki2009non}, since they can be regarded as a non-local version of the Fisher-KPP model~\eqref{FKPP}-\eqref{def:RFKPP}. However, note that in the literature the term non-local Fisher-KPP equation is sometimes used, by extension, to refer to non-local PDE models of form~\eqref{nonlocalFKPP} as well. \label{footnote:mut-sel}} 
\begin{equation}\label{nonlocalFKPP}
\begin{cases}
\displaystyle{\partial_t n = \bar{D} \nabla_{\bm y}^2 n + n R({\bm {y}},\rho)\,,\quad {\bm y} \in \mathcal{Y} \ ,  }\\[5pt]
\displaystyle{\rho(t) := \int_{{\cal {Y}}} n(t,{\bm y}) \, {\rm d}{\bm y}\,,}
\end{cases}
\end{equation}
where now $\rho(t)$ is the (total) population size at time $t$. Models of this type have been widely used in theoretical studies into the evolutionary dynamics of phenotype-structured populations, for different biological contexts~\citep{perthame2006transport}. Focusing on cell populations, a possible form of the fitness function $R({\bm {y}},\rho)$ that has been employed, e.g. by~\cite{almeida2019evolution,lorenzi2016tracking}, to investigate the evolutionary dynamics of cancer cells in spatially homogeneous scenarios is   
\begin{equation}\label{def:REx1}
R({\bm y},\rho) := r({\bm y}) - \kappa \rho \, .
\end{equation}
The term $r({\bm y})$ in~\eqref{def:REx1} can be regarded as the net per capita growth rate (i.e. the difference between the rate of proliferation and the rate of death under natural selection) of population members in the phenotypic state ${\bm y}$. Hence the maximum points of this function, which coincide with the maximum points of the fitness function $R({\bm y},\rho)$, correspond to the fitness peaks (i.e. the peaks of the phenotypic fitness landscape of the population)~\citep{diekmann2005dynamics,lorenzi2020asymptotic}. For example, in scenarios where there is one single fitness peak, a possible simple definition of the function $r$ is  
\begin{equation}\label{def:pEx1}
r({\bm y}) := \gamma - \left|{\bm y} - {\bm \varphi}\right|^2 \, .
\end{equation}
Here the point ${\bm \varphi} \in \mathcal{Y}$ models the phenotypic state corresponding to the fitness peak (i.e. the fittest phenotype) and the parameter $\gamma \in \mathbb{R}^+_{*}$ is linked to the maximum fitness (i.e. the fitness of the fittest phenotype). Moreover, the saturating term $- \kappa \rho$ in~\eqref{def:REx1} models the limitations on population growth imposed by carrying capacity constraints, and the parameter $\kappa \in \mathbb{R}^+_*$ is inversely related to the carrying capacity of the population. Of course, when modelling the evolutionary dynamics of cell populations, alternatives to~\eqref{def:REx1} can also be considered for the fitness function, as done, for instance, by~\cite{lorz2013populational,chisholm2015emergence,guilberteau2023integrative}. 

\begin{figure}
    \centering
    \includegraphics[width=\linewidth]{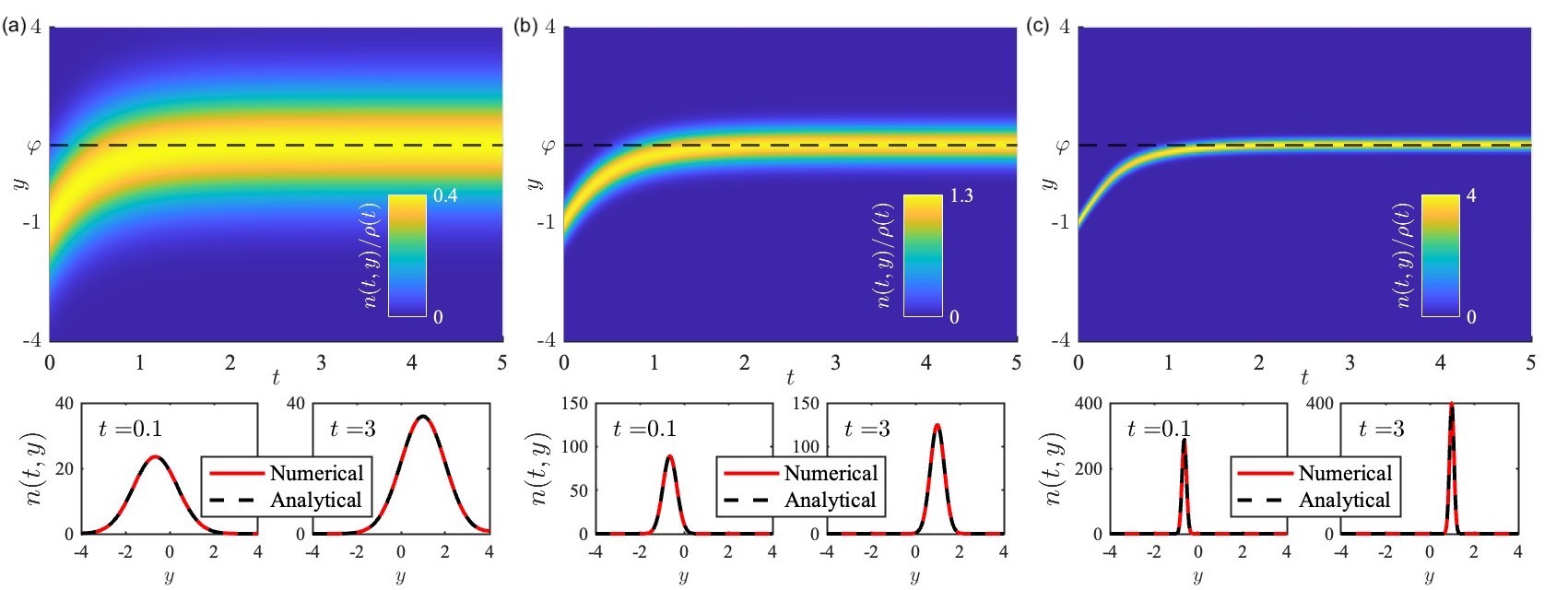}
\caption{{\bf Illustration of Gaussian-like solutions and concentration phenomena from a non-local PDE model of evolutionary dynamics.} Illustrative output of the non-local PDE model of evolutionary dynamics~\eqref{nonlocalFKPP}-\eqref{def:pEx1}, with ${\bm y} \equiv y \in [-5,5]$, subject to zero-flux boundary conditions. In particular, Gaussian-like solutions and concentration phenomena are shown for the rescaled non-local PDE~\eqref{nonlocal-FisherKPP:rescaled}, under definitions~\eqref{def:REx1},\eqref{def:pEx1} (with $\kappa=0.1$, $\gamma=10$, and $\varphi=1$), subject to initial condition~\eqref{eq:ICNLFKPPrev} (with $\rho^0=60$ and $\bar{y}^0=-1$). Simulations are for (a) $\varepsilon=1$, (b) $\varepsilon=0.1$, and (c) $\varepsilon=0.01$. Top panels display the dynamics of the rescaled phenotype density $n(t,y)/\rho(t)$; bottom panels display the phenotype density $n(t,y)$ (red solid line) at select times. We compare the numerical solution with the analytically predicted one (black dashed lines) -- i.e. the Gaussian-like solution~\eqref{eq:SOLNLFKPPreveps} with mean $\bar{y}_\varepsilon(t)$ and weight $\rho_\varepsilon(t)$ obtained by solving numerically the Cauchy problem~\eqref{eq:NLFKPPodesreveps}. We see the solution to be of a Gaussian form, which becomes concentrated as a weighted increasingly sharp Gaussian as $\varepsilon$ gets smaller, corroborating the emergence of concentration phenomena of type~\eqref{eq:concphen}. We see also that the mean or prevailing phenotype $\bar{y}(t)$ converges to $y=\varphi$ (black dashed lines, top panels) as $t \to \infty$, corresponding to selection of the fittest trait.}  \label{fig:gaussiansol}
\end{figure}

As illustrated by the plots in Figure~\ref{fig:gaussiansol}, when $p=1$, and thus ${\bm y} \equiv y$ and ${\bm \varphi} \equiv \varphi$, non-local PDE models of form~\eqref{nonlocalFKPP} complemented with definitions~\eqref{def:REx1},\eqref{def:pEx1} admit Gaussian-like solutions with a variance proportional to $\bar{D}$ and a mean, $\bar{y}(t)$, which represents the mean or prevailing phenotype in the population at time $t$, that converges to $\varphi$ as $t \to \infty$~\citep{almeida2019evolution,ardavseva2020evolutionary,chisholm2016evolutionary}. This provides a possible mathematical formalisation of two key facts in evolutionary biology: the rate at which individuals undergo phenotypic changes, $\bar{D}$, impacts on the phenotypic variance (i.e. the level of phenotypic heterogeneity) of the population; and the mean or prevailing phenotype in the population will ultimately be the fittest phenotype $\varphi$. Moreover, as discussed in detail in Section~\ref{sec:analysis:CPNPDE} and as illustrated again by Figure~\ref{fig:gaussiansol}, appropriately rescaled versions of these models can exhibit concentration phenomena -- i.e. their solutions can become concentrated as weighted infinitely sharp Gaussians, that is, weighted Dirac masses -- formalising in mathematical terms the idea that the population shows only one trait, and thus remains monomorphic over time~\citep{diekmann2005dynamics,perthame2006transport}. We note that other forms of explicit and semi-explicit solutions to variants of the non-local PDE~\eqref{nonlocalFKPP} have also been obtained, e.g. see~\cite{alfaro2014explicit,alfaro2017replicator}, and cases where there are multiple fitness peaks have also been analytically investigated, e.g. see~\cite{alfaro2019evolutionary,lorenzi2020asymptotic}.

Finally, for a spatially heterogeneous environment -- for example a  population surrounded by non-uniformly distributed abiotic factors (nutrients, therapeutic agents, etc) that affects proliferation and death rates -- a natural extension of \psipde models of form (\ref{structuredFKPP}) is 
\begin{equation}\label{nonlocal-FisherKPPhetenv}
\begin{cases}
\displaystyle{\partial_t n = D({\bm y}) \nabla_{\bm x}^2 n + \bar{D} \nabla_{\bm y}^2 n + n R({\bm y},\rho,{\bm S}) \ , \quad {\bm x} \in \mathcal{X} \ , \; {\bm y} \in \mathcal{Y} \ ,  }\\[5pt]
\displaystyle{\rho(t,{\bm x}) := \int_{{\cal {Y}}} n(t,{\bm x},{\bm y}) \, {\rm d}{\bm y} \ .}
\end{cases}
\end{equation}
Here the vector ${\bm S}(t,{\bm x})=\left(S_1(t,{\bm x}), \ldots, S_N(t,{\bm x})\right)$ represents the concentrations at position ${\bm x}$ at time $t$ of $N$ different abiotic factors, the dynamics of which are governed by appropriate evolution equations coupled with the PS-PDE~\eqref{nonlocal-FisherKPPhetenv}. In this more general context, definition~\eqref{def:REx1} and definition~\eqref{def:pEx1} can be modified, respectively, as
\begin{equation}\label{def:REx1hetenv}
R({\bm y},\rho,{\bm S}) := r({\bm y},{\bm S}) - \kappa \rho
\end{equation}
and
\begin{equation}\label{def:pEx1hetenv}
r({\bm y},{\bm S}) := g({\bm S}) - \left|{\bm y} - f({\bm S})\right|^2 \, ,
\end{equation}
where $f : \mathbb{R}^N \to \mathcal{Y}$ models the fittest phenotype under the selective pressure of the local environmental conditions, which are determined by the concentrations of the different abiotic factors, and $g : \mathbb{R}^N \to \mathbb{R}^+_{*}$ is related to the corresponding maximum fitness. 

Starting from~\cite{lorz2015modeling}, where a \psipde model of form (\ref{nonlocal-FisherKPPhetenv}) with $D({\bm y}) \equiv 0$ and $\bar D = 0$ was proposed to study the emergence of cell phenotypic heterogeneity in avascular tumours, models that can be regarded as variations on the theme of~\eqref{nonlocal-FisherKPPhetenv} have found application in a number of theoretical studies on the eco-evolutionary dynamics of cancer cells in avascular and vascularised tumours -- e.g. see~\cite{celora2023spatio,chiari2023hypoxia,chiari2023hypoxiab,fiandaca2021mathematical,lorenzi2018role,villa2021modeling,villa2021evolutionary}. Variants of~\eqref{nonlocal-FisherKPPhetenv} with $D({\bm y})\equiv D$ have also received attention from the mathematical community as a mean to investigate analytically evolutionary dynamics in phenotype-structured populations exposed to shifting environments~(\citealt{alfaro2013travelling,alfaro2017effect,berestycki2013traveling}) and spatially periodic environments~\citep{boutillon2024reaction}. 

\subsection{Pressure-based movement models}

\begin{figure}
    \centering
    \includegraphics[width=1\linewidth]{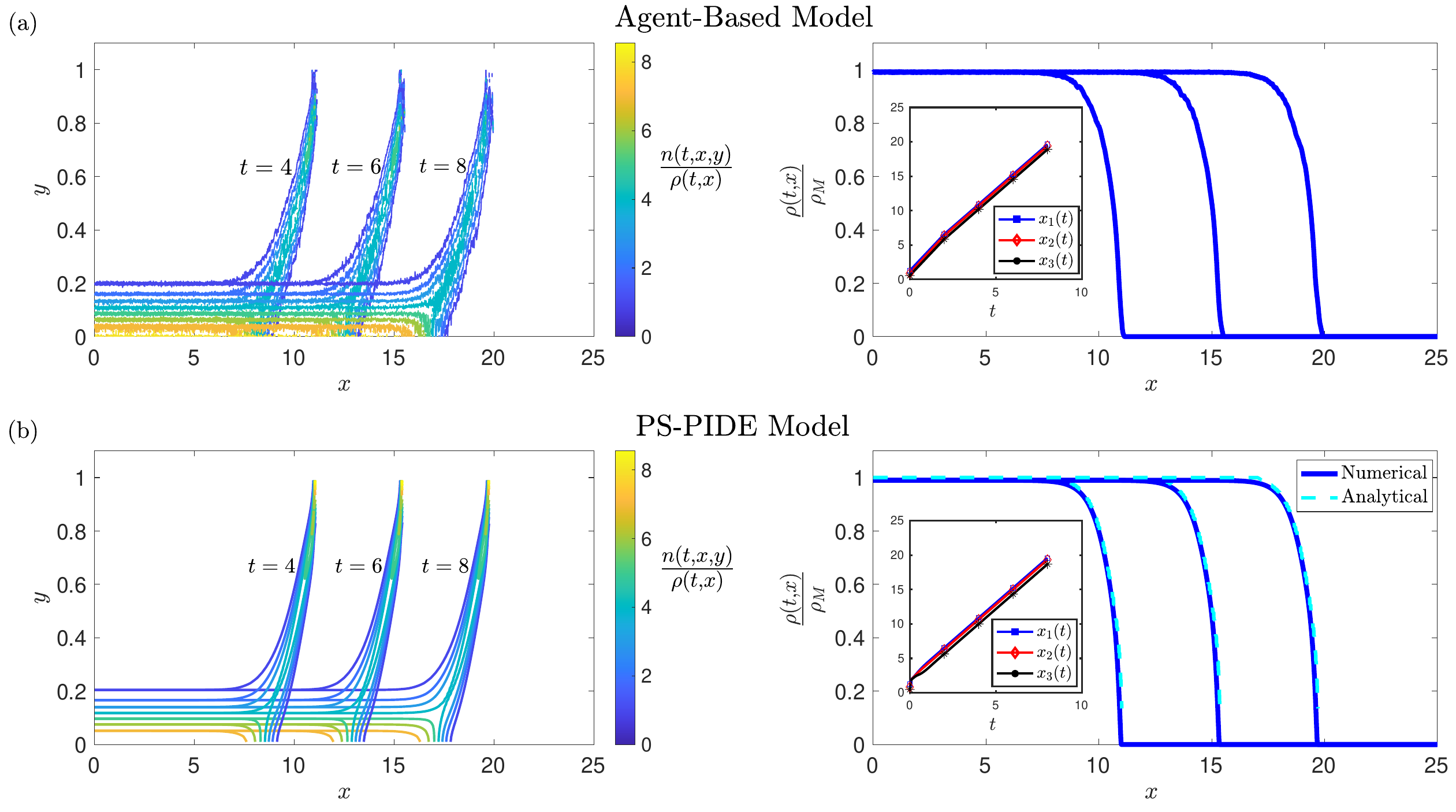}
    \caption{{\bf Illustration of phenotypic structuring across travelling waves from a \psipde model of pressure-based movement.} Illustrative outputs of (b) a PS-PDE model of pressure-based movement of form~\eqref{structuredPM} and (a) a corresponding ABM of the type presented in Section~\ref{sec:IBmodel}. In particular, phenotypic structuring across travelling waves is shown for the rescaled PS-PDE~\eqref{PSPIDEP:rescaled} with $\varepsilon=0.01$ and the corresponding rescaled ABM. Here, we consider one dimension in each of physical and phenotype space, such that ${\bm x} \equiv x \in[0,25]$ and ${\bm y} \equiv y \in[0,1]$. Zero-flux conditions are imposed at the boundaries. These results are for a go-or-grow type scenario, where the phenotypic state $y\in[0,1]$ is such that $y$ close to $0$ corresponds to high proliferation rate and low mobility coefficient, while $y$ close to 1 corresponds to low proliferation rate and high mobility coefficient. We see an invasion process in which a travelling wave of cells progresses, and there is a good agreement between the results of numerical simulations of the ABM, numerical solutions of the PS-PDE model, and analytical predictions (right panels). Cells with a higher mobility coefficient are concentrated to the front of the wave, and cells with a higher proliferation rate dominate the rear (left panels). The insets of the right panels display the plots of the points $x_1(t)$, $x_2(t)$, and $x_3(t)$ such that $\rho(t,x_1(t))= 0.2 \rho_M$, $\rho(t,x_2(t))= 0.5 \rho_M$, and $\rho(t,x_3(t))= 0.8 \rho_M$, where $\rho_M \in \mathbb{R}^+_*$ is the maximum value of the cell density. Notably, we observe the evolution of $x_1(t)$, $x_2(t)$, and $x_3(t)$ towards straight lines of approximately the same slope, and the slope agrees with analytical predictions on the wave speed. This indicates that there is evolution towards a travelling wave with a wave speed consistent with analytical predictions. For full details of the models and parametrisation, we refer to~\cite{macfarlane2022individual}: this figure corresponds to Figure~2 of~\cite{macfarlane2022individual}. }
    \label{pressurebasedfigure}
\end{figure}

Diffusion-based models of cell movement implicitly rely on the assumption that cell motion is undirected. However, cells, especially eukaryotic cells, can have a tendency to move towards regions in which they are less compressed~\citep{byrne2003modelling,byrne1997free}. This aspect of cell movement, which is central to a variety of physiological and pathological processes -- encompassing tissue development, wound healing, and avascular tumour growth -- can be captured by pressure-based models~\citep{roose2007mathematical}. Building on the ideas presented in the seminal paper by \cite{greenspan1976growth} and subsequent papers~\citep{ambrosi2002closure,bresch2010computational,byrne2009individual,byrne2003modelling,byrne1997free,ciarletta2011radial,lowengrub2009nonlinear,preziosi2009multiphase,sherratt2001new}, the basic tenet of these models is that cell movement is described via an advective term with an advective velocity inversely proportional to the gradient of the cellular pressure, $P(t,{\bm x})$. This leads to a PDE of the following form, which can be obtained from (\ref{dar}) by taking $D_{\bm x} \equiv 0$, ${\bm A_{\bm x}} := - \mu \nabla_{\bm x} P$, and $F := \rho R(P)$:
\begin{equation}\label{PM}
\begin{cases}
\displaystyle{\partial_t \rho = \nabla_{\bm x} \cdot \left[\mu \ \rho \nabla_{\bm x} P\right] + \rho R(P)\, , \quad {\bm x} \in \mathcal{X} \ ,  }\\[5pt]
\displaystyle{P(t,{\bm x}) := \Pi[\rho](t,{\bm x}) \ .}
\end{cases}
\end{equation}
In~\eqref{PM}, the function $\Pi(\rho)$ represents a constitutive law (i.e. a barotropic relation) for the cellular pressure as a function of the cell density (i.e. the cell volume fraction), which is needed to close the PDE for $\rho$~\citep{ambrosi2002closure}. In analogy with Darcy's law for fluid flow in porous media~\citep{darcy1856fontaines}, the parameter $\mu \in \mathbb{R}^+_*$ is usually referred to as the cell mobility coefficient. This is inversely proportional to the permeability of the medium in which the cells are embedded (\eg the extracellular matrix) and depends on the cells' morphological and mechanical properties (e.g. cell elongation and nucleus deformability). Here $\mu$ is taken to be constant but it may also be a function of ${\bm x}$ and $t$ due to the presence of inhomogeneities and anisotropicities, as well as the occurrence of temporal changes, in the embedding medium that surrounds the cells. 

The growth rate of the cell density, $R$, depends on $P$ to incorporate pressure-dependent inhibition of growth (i.e. cessation of cell division when cellular pressure becomes too high)~\citep{drasdo2012modeling,ranft2010fluidization}. Note that, if the growth of the cell population was not only pressure-regulated but also nutrient-limited, then the growth rate would also depend on the nutrient concentration, and an additional PDE for the nutrient concentration would be introduced. Models of form~\eqref{PM} -- and related variants for multiple cell populations -- have  drawn interest from mathematicians and physicists for their ability to recapitulate key aspects of tumour and tissue growth, and also for exhibiting travelling waves which display interesting features -- e.g. see~\citep{bertsch2015travelling,carrillo2024multipop,chaplain2020bridging,lorenzi2016interfaces,tang2014composite}.

Note also that when one takes $\Pi(\rho):=K \rho$, where $K \in \mathbb{R}^+_*$ is a scale factor (which for simplicity can be set as $1$), the transport term in the PDE~\eqref{PM} models more specifically the tendency of cells to disperse to avoid crowding (i.e. to move down the gradient of the cell density $\rho$)~\citep{chaplain2006mathematical}, while the reaction term takes into account density-dependent inhibition of growth~\citep{lieberman1981density}. Moreover, as discussed by~\cite{perthame2014hele}, when $\Pi(\rho):= K_{\gamma} \rho^{\gamma}$, where the parameter $\gamma \in \mathbb{R}^+_*$ with $\gamma>1$ provides a measure of the stiffness of the pressure law and $K_{\gamma} \in \mathbb{R}^+_*$ is a scale factor, the PDE~\eqref{PM} takes the form of a porous medium-type equation. In this case, considering the asymptotic regime $\gamma \to \infty$\footnote{Since the parameter $\gamma$ provides a measure of the stiffness of the pressure law $\Pi(\rho)$, the limiting regime $\gamma \to \infty$ is usually referred to as `incompressible limit', because it corresponds to mathematically approximating cells as an incompressible fluid.}, from the PDE~\eqref{PM} one can derive a free-boundary problem of Hele-Shaw type, as demonstrated by~\cite{perthame2014hele}. This forms a bridge between PDE models of form~\eqref{PM} and mathematical models formulated as free-boundary problems, which have also been widely employed~\citep{friedman2015free}. This is also an aspect that has received increasing attention from the mathematical community -- see e.g.~\cite{bubba2020hele,david2024incompressible,kim2016free,mellet2017hele}. 

Generalisations of PDE models of the form~\eqref{PM} into \psipde models of form
\begin{equation}\label{structuredPM}
\begin{cases}
\displaystyle{\partial_t n = \nabla_{\bm x} \cdot \left[\mu({\bm y}) n \nabla_{\bm x} P\right] + \bar{D} \nabla_{\bm y}^2 n +  n R({\bm y},P)\, , \quad {\bm x} \in \mathcal{X} \ , \; {\bm y} \in \mathcal{Y} \ ,  }\\[5pt]
\displaystyle{P(t,{\bm x}) := \Pi[\rho](t,{\bm x}), \;\; \rho(t,{\bm x}) := \int_{{\cal {Y}}} n(t,{\bm x},{\bm y}) \, {\rm d}{\bm y} \ }
\end{cases}
\end{equation}
have been proposed for the situation in which both the rate of proliferation and the mobility coefficient of the cells depend on the phenotypic state ${\bm y}$. Specifically, a \psipde model of this form with $\Pi(\rho):= \rho$ has been proposed by~\cite{lorenzi2022invasion}, while different possible definitions of $\Pi(\rho)$ have been considered by~\cite{macfarlane2022individual}. PS-PDE models of type~\eqref{structuredPM} can be obtained from~\eqref{general} by taking $D_{\bm x} \equiv 0$ and ${\bm A_{\bm x}} := - \mu({\bm y}) \nabla_{\bm x} P$ along with $F := n R({\bm y},P)$, ${\bm A}_{\bm y} \equiv 0$, and $D_{\bm y} \equiv \bar D$. Here the function $\mu({\bm y})$ is the mobility coefficient of cells in the phenotypic state ${\bm y}$. Furthermore, the function $R({\bm y},P)$ models the net proliferation rate of cells in the phenotypic state ${\bm y}$ under the local environmental conditions corresponding to the cellular pressure $P$, and can thus be regarded again as a fitness function. This can be defined along the lines of~\eqref{def:REx1} as  
\begin{equation}\label{def:REP}
R({\bm y},P) := r({\bm y}) - \kappa P \, ,
\end{equation}
where the parameter $\kappa \in \mathbb{R}^+_*$ is inversely related to the critical value of the cellular pressure above which cessation of cell division occurs. 

Considering `go-or-grow’ type scenarios in which fast proliferating cells are less migratory (i.e. they have a lower mobility coefficient) and vice versa, \cite{lorenzi2022invasion,macfarlane2022individual} have shown that \psipdes of form~\eqref{structuredPM} can produce phenotype-structured travelling waves whereby fast proliferating cells make up the bulk of the population in the rear of the wave, while highly migratory cells drive invasion at the edge of the wave front (see Figure~\ref{pressurebasedfigure}). Focusing on the case where all cells in the population have the same mobility coefficient regardless of their phenotypic state (i.e. when $\mu({\bm y}) \equiv \mu$), connections between \psipde models of this type and free boundary problems of Hele-Shaw type have been explored by~\cite{david2023phenotypic}, while related models have been employed to investigate the emergence of resistance to chemotherapy alongside tumour growth by~\cite{cho2018modeling}. Moreover, extensions of~\eqref{structuredPM} have been considered by~\cite{fiandaca2022phenotype} to model the growth of avascular tumours, taking into account tumour necrosis and tumour-microenvironment interactions.

\subsection{Taxis-based movement models}

\begin{figure}[t!]
    \centering
    \includegraphics[width=\textwidth]{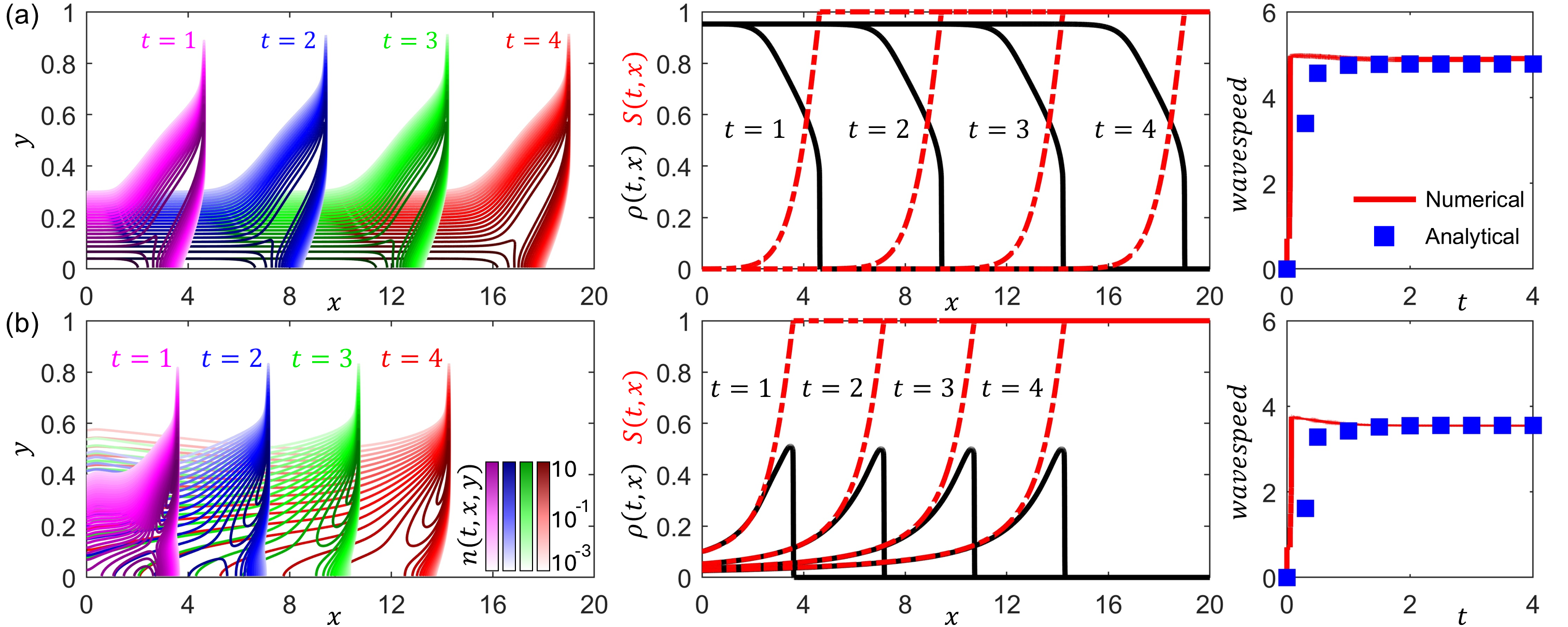}
    \caption{
    {\bf Illustration of phenotypic structuring across travelling waves from a \psipde model of chemotaxis-based movement.} Illustrative outputs of a PS-PDE model of chemotaxis-based movement of form~\eqref{structuredKS}. In particular, phenotypic structuring across travelling waves is shown for the rescaled \psipde~\eqref{PSPIDES:rescaled} with $\varepsilon=0.01$ coupled with an appropriate  evolution equation for the attractant concentration, $S$. Here, we consider one dimension in each of physical and phenotype space, such that ${\bm x} \equiv x \in[0,20]$ and ${\bm y} \equiv y \in[0,1]$. Zero-flux conditions are imposed at the boundaries. (a) Chemotaxis-proliferation trade-off, where the phenotypic state $y\in[0,1]$ is such that $y$ close to $0$ corresponds to high proliferation rate and low chemotactic sensitivity, while $y$ close to 1 corresponds to low  proliferation rate and high chemotactic sensitivity. All cells equally degrade the attractant. We see an invasion process in which as the wave progresses, nutrient is degraded (left panel); the most efficient chemotaxers are concentrated to the front of the wave, and strong proliferators dominate the rear (middle panel); the numerically estimated wavespeed agrees with analytical predictions (right panel). (b) Chemotaxis-proliferation trade-off for a nutrient-like attractant, i.e. an attractant that has an impact on the cell  proliferation rate. The phenotypic state $y\in[0,1]$ is again such that $y$ close to $0$ corresponds to high proliferation rate but low chemotactic sensitivity, while $y$ close to 1 corresponds to low proliferation rate and high chemotactic sensitivity. We see a pulse waveform for the population (left panel); as in (a), strongly (weakly) chemotactic cells are concentrated to the front (rear) of the wave (middle panel); numerically estimated wavespeed agrees with analytical predictions (right panel). For full details of the model and parametrisation, we refer to \cite{lorenzi2022invasion}: panel (a) corresponds to Figure~3A and panel (b) corresponds to Figure~5B of \cite{lorenzi2022invasion}, respectively.}
    \label{chemotaxisfigure}
\end{figure}

In processes that include morphogenesis, wound healing, and inflammation, cell populations need to be correctly positioned. Taxis-based\footnote{The word taxis stems from the ancient Greek $\tau\alpha\xi\iota\sigma$, meaning `to steer’, and was adopted by Wiliam Pfeffer to describe a hypothetical steering of microorganisms, following observations of their reorganisation according to chemicals and light \citep{pfeffer1884}.} models form a third major class of movement models, describing situations in which a cell (or organism) is directed to move in a particular direction. Chemotaxis, the directed movement of cells up or down gradients of (usually soluble) molecules, has been the most intensely studied form of taxis behaviour, but a wide variety of others have also been investigated, including movement in response to adhesion gradients (haptotaxis, \citealt{carter1965principles}), stiffness (durotaxis, \citealt{lo2000cell}), and electric fields (galvanotaxis, \citealt{erickson1984embryonic}). 

The most well-known PDE model for describing chemotactic movements is the Keller-Segel system~\citep{keller1970initiation,keller1971traveling,keller1971model,patlak1953random}. This model has received significant attention as a model for chemotaxis \citep{horstmann2003from,tindall2008overview,hillen2009user,bellomo2015toward,painter2019mathematical,arumugam2021keller}: from the modelling community for its ability to recapitulate macroscopic dynamics of chemotactic populations, including travelling waves and self-organisation, and from the mathematical community for its intricate analytical properties. Moreover, it has provided an important base for describing other forms of taxis behaviour, with its adaption to describe haptotaxis a particularly prominent example~\citep{wang2020review,sfakianakis2020mathematical}.  
In its standard form, (chemo)taxis is described via an advective term with an advective velocity proportional to the gradient of a (chemical) signal, $S(t,{\bm x})$. Along with standard choices of an additional linear diffusion component (i.e. similarly to what done in~\eqref{FKPP}) and a growth term, one sets $D_{\bm x} \equiv D$, ${\bm A_{\bm x}} := \chi \nabla_{\bm x} S$, and $F := \rho R(\rho,S)$ in (\ref{dar}) to obtain
\begin{equation} \label{KS}
	\partial_t \rho = \nabla_{\bm x} \cdot \left[  D \nabla_{\bm x} \rho - \chi \rho \nabla_{\bm x} S\right]+ \rho R(\rho,S)\,, \quad {\bm x} \in \mathcal{X} \ .  
\end{equation}
The parameter $\chi \in \mathbb{R}^+_*$ is usually referred to as the chemotactic sensitivity -- here it is taken to be a constant, but it can be chosen to depend on the signal $S$ and/or the density $\rho$~\citep{hillen2009user}. Note that the growth rate may depend on the signal, as it is the case in the PDE~\eqref{KS}, for example if this represents a nutrient that fuels proliferation. In many models, also the dynamics of $S(t,{\bm x})$ will be governed by some PDE, leading to a system of coupled PDEs. Such couplings are crucial in the context of pattern formation phenomena, where feedback between chemotaxis and signal production leads to self organisation~\citep{keller1970initiation} and travelling wave dynamics, where signal degradation by cells can generate propagating waves~\citep{keller1971traveling}.   

An extension of (\ref{KS}) to a \psipde model for chemotaxis has been proposed by \cite{lorenzi2022trade}, 
focusing on the case where phenotype regulates both proliferation and chemotaxis. One can arrive at this form of model through extending to a phenotype density $n(t,{\bm x},{\bm y})$ and -- with the same earlier choice of linear diffusion in phenotype (\ie taking ${\bm A}_{\bm y} \equiv 0$ and $D_{\bm y} \equiv \bar D$ in~\eqref{general}) -- obtaining the following \psipde formulation of~\eqref{KS}:
\begin{equation}\label{structuredKS}
\begin{cases}
\displaystyle{\partial_t n = \nabla_{\bm x} \cdot \left[D \nabla_{\bm x} n - \chi({\bm y}) n \nabla_{\bm x} S\right] + \bar{D} \nabla_{\bm y}^2 n +  n R({\bm y},\rho,S)\, , \; {\bm x} \in \mathcal{X} \ , \; {\bm y} \in \mathcal{Y} \ , }\\[5pt]
\displaystyle{\rho(t,{\bm x}) := \int_{{\cal {Y}}} n(t,{\bm x},{\bm y}) \, {\rm d}{\bm y} \ .}
\end{cases}
\end{equation}
In the above model, each of the chemotactic sensitivity and growth rate now depend on the phenotypic state -- i.e. the function $\chi({\bm y})$ models the chemotactic sensitivity of cells in the phenotypic state ${\bm y}$, while the function $R({\bm y},\rho,S)$ is the net proliferation rate (i.e. the fitness) of cells in the phenotypic state ${\bm y}$ under the local environmental conditions corresponding to the cell density $\rho$ and the signal $S$, which can for example be defined via~\eqref{def:REx1hetenv} with ${\bm S} \equiv S$. Experimental studies with microfluidic T-mazes have revealed that bacterial populations can be structured according to their chemotactic sensitivity (\citealt{salek2019bacterial}, see Figure \ref{schematicfigure} (b)), while other studies have suggested that resource allocation can lead to a trade-off between investment in motility and growth \citep{ni2020growth}. Consequently, the case of a `proliferation-chemotaxis’ trade-off -- that is, faster proliferators are less chemotactic (i.e. they have a lower chemotactic sensitivity) and vice versa -- was considered by~\cite{lorenzi2022trade}, who demonstrated that \psipdes of form~\eqref{structuredKS} coupled with an appropriate evolution equation for $S$ can form travelling waves that exhibit phenotypic structuring (see Figure~\ref{chemotaxisfigure}).

Of course, one can consider a number of further variations. For instance, letting $S$ model the density of the extracellular matrix, \psipdes of form~\eqref{structuredKS} have been employed to describe cancer cell dynamics in phenotype-structured haptotaxis models of tumour invasion~\citep{fiandaca2022phenotype,lorenzi2023derivation}. Related PS-PDE models of cell migration, where $S$ is replaced by a vector ${\bm S}$, the components of which model the concentrations of different diffusible molecules, have been considered by~\cite{hodgkinson2018signal}. Variants of PS-PDEs of form~\eqref{structuredKS}, wherein a phenotypic structure is included in the growth term but not in the term modelling cellular taxis, have been used in models for cell migration~\citep{domschke2017structured} and tumour cell invasion~\citep{engwer2017structured,hodgkinson2018computational,hodgkinson2019spatio}. 

PS-PDE models closely related to (\ref{structuredKS}) have also been elegantly deployed alongside experiments, to investigate how phenotype structuring drives collective behaviour within travelling {\em E. coli} bands \citep{fu2018spatial,mattingly2022collective,phan2024direct}. In these models, the structuring describes a bacterium's tumbling bias and manifests in phenotype dependency for both cell diffusion,  $D_{\bm x} := D({\bm y})$, and chemotactic sensitivity, $\chi({\bm y})$. These models assume that bacteria do not alter phenotype within their lifetime (i.e. ${\bm A}_{\bm y} \equiv 0$ and $D_{\bm y} \equiv 0$ in~\eqref{general}), but can as a result of division, an aspect that is included in the model via a kernel that stipulates how division of a parent cell with one phenotype leads to progeny cells with other phenotypes. This leads to a non-local form of the term $F$ in~\eqref{general} that depends on the phenotype structuring. Direct comparisons of this model with data have revealed the limitations of `traditional' non-structured models and gained insights into how an evolving structuring may allow populations to efficiently colonise heterogeneous environments.

\section{Mathematical tools and techniques} \label{techniques} 

In this section, we provide a concise overview of mathematical tools and techniques that have been developed and deployed to: derive PS-PDE models of form~\eqref{general} from agent-based models for the spatial spread and evolutionary dynamics of phenotype-structured cell populations (Section~\ref{sec:tools:derivation}); analyse the qualitative properties of the solutions to such PS-PDE models (Section~\ref{sec:tools:analysis}); and construct numerical solutions of these models (Section~\ref{sec:tools:numerics}). 

For simplicity, we now restrict to one-dimensional physical and phenotypic domains -- i.e. the case where ${\bm x} \equiv x \in \mathcal{X} \subseteq \mathbb{R}$ and ${\bm y} \equiv y  \in \mathcal{Y} \subseteq \mathbb{R}$. Throughout this section, we will let $\varepsilon \in  \mathbb{R}^+_*$ be a small parameter, and use the notation $\delta_{a}(y)$ for the Dirac delta centred at $y=a$.


\subsection{Tools and techniques to derive PS-PDE models from agent-based models}
\label{sec:tools:derivation}
Instead of defining a PS-PDE model of type~\eqref{general} on the basis of population-scale phenomenological assumptions, it can be desirable to first postulate an ABM that tracks the spatial and evolutionary dynamics of individual cells, and then employ limiting procedures to derive the corresponding PS-PDE model~\citep{drasdo2005coarse}. In this way, the population-level terms that comprise the PS-PDE can be formally linked to explicit assumptions for the cellular processes that drive the dynamics at the single-cell level. 

A range of approaches have been advanced and adopted in recent decades for transitioning between ABMs and PS-PDE models which describe simultaneously the spatial spread and evolutionary dynamics of phenotype-structured populations. These include probabilistic methods to derive PS-PDE models as the limit of corresponding off-lattice ABMs when the number of individuals in the population tends to infinity (see e.g. ~\citealt{andrade2019local,champagnat2007invasion,fontbona2015non,leman2016convergence}), and formal limiting procedures to derive PS-PDE models from on-lattice ABMs when the lattice parameters tend to zero (see e.g. ~\citealt{macfarlane2022individual,lorenzi2023derivation}). As a further note, structuring variables have also been incorporated into `mesoscale' kinetic equations, which can also be scaled to a PDE form (see e.g. \citealt{erban2004individual,engwer2015glioma,lorenzi2024phenotype}); we are, however, unaware of any such derivations that lead explicitly to the PS-PDE form (\ref{general}).

In the spirit of a tutorial-style example, here we formally derive a PS-PDE model of form~\eqref{general}, starting from an ABM for the spatial spread and evolutionary dynamics of phenotype-structured cell populations (see Section~\ref{sec:derfromIBmodel}). In this model, single cells undergo spatial movement, phenotypic changes, and proliferation and death according to a set of rules which correspond to a discrete-time branching random walk over a two-dimensional lattice~\citep{hughes1996random}, which represents one-dimensional physical and phenotypic domains (see Section~\ref{sec:IBmodel}). Note that here we will exemplify key ideas considering the case of unbounded domains, but both the governing rules for the dynamics of single cells in the ABM and the procedure employed to derive the corresponding PS-PDE can easily be adapted to the case of bounded domains. 

\subsubsection{An agent-based model for spatial spread and evolutionary dynamics of phenotype-structured cell populations}
\label{sec:IBmodel}
\paragraph{Setup of the model and notation}
The time variable $t\in\mathbb{R}^+$ and the space variable $x\in\mathbb{R}$ are discretised, respectively, as $t_k=k \,  \tau$ and $x_i = i h_x$, with $k\in \mathbb{N}_0 := \mathbb{N} \cup \{0\}$, $\tau \in \mathbb{R}^+_*$, $i\in\mathbb{Z}$, and $h_x \in \mathbb{R}^+_*$.  Moreover, the phenotype variable $y \in \mathbb{R}$ is discretised via $y_j=j h_y$, with $j\in\mathbb{Z}$ and $h_y \in \mathbb{R}^+_*$. Here, $\tau$, $h_x$, and $h_y$ are the time-step, space-step, and phenotype-step, respectively.

Each individual cell is represented as an agent that occupies a position on the lattice $\{x_i\}_{i\in\mathbb{Z}}\times\{y_j\}_{j\in\mathbb{Z}}$ and we introduce the dependent variable $N_{i,j}^k\in\mathbb{N}_0$ to model the number of cells in the phenotypic state $y_j$ at position $x_i$ at time $t_k$. The cell phenotype density and the corresponding cell density are then defined, respectively, as
\begin{equation}\label{eq:nNIB}
n(t_k,x_i,y_j) \equiv n^k_{i,j} :=\frac{N_{i,j}^k}{h_x h_y}, \qquad \rho(t_k,x_i) \equiv \rho^k_i :=h_y \sum_j n_{i,j}^k.
\end{equation}
The modelling strategies adopted here to describe spatial movement, phenotypic changes, and proliferation and death of individual cells are summarised below. 

\paragraph{Modelling spatial movement of individual cells}
Cell movement is modelled as a random walk along the spatial dimension. For a focal cell in the phenotypic state $y_j$ at spatial position $x_i$ at time $t_k$, the probability of moving left (i.e. to spatial position $x_{i-1}$) is $P^L_x(t_k,x_i,y_j)$, the probability of moving right (i.e.  to spatial position $x_{i+1}$) is $P^R_x(t_k,x_i,y_j)$, and the probability of not moving (i.e. remaining stationary at position $x_i$) is then $P^S_x(t_k,x_i,y_j) = 1 - P^R_x(t_k,x_i,y_j) - P^L_x(t_k,x_i,y_j)$. Here, $P^L_x$ and $P^R_x$ are non-negative real functions such that $P^L_x(t_k,x_i,y_j) + P^R_x(t_k,x_i,y_j) \leq 1$ for all $(t_k,x_i,y_j) \in \mathbb{R}^+ \times \mathbb{R} \times \mathbb{R}$.

\paragraph{Modelling phenotypic changes of individual cells}
Phenotypic changes are incorporated into the model by allowing cells to update their phenotypic state according to a random walk along the phenotypic dimension. For a focal cell in the phenotypic state $y_j$ at spatial position $x_i$ at time $t_k$, the probability of entering the phenotypic state $y_{j-1}$ is $P^L_y(t_k,x_i,y_j)$, the probability of entering the phenotypic state $y_{j+1}$ is $P^R_y(t_k,x_i,y_j)$, and the probability of not undergoing phenotypic changes (i.e. remaining in the phenotypic state $y_j$) is $P^S_y(t_k,x_i,y_j) = 1 - P^R_y(t_k,x_i,y_j) - P^L_y(t_k,x_i,y_j)$. To reduce the amount of calculations in the derivation of the corresponding PS-PDE model, we focus on the case where cells in the population undergo phenotypic changes with a constant probability, and thus we assume $P^L_y(t_k,x_i,y_j)\equiv p^L_y\in \mathbb{R}^+$ and $P^R_y(t_k,x_i,y_j)\equiv p^R_y\in \mathbb{R}^+$, which imply $P^S_y(t_k,x_i,y_j) \equiv p^S_y = 1 - p^R_y - p^L_y$, with the parameters $p^L_y$ and $p^R_y$ such that $p^L_y + p^R_y \leq 1$.

\paragraph{Modelling proliferation and death of individual cells}
Cell proliferation is modelled by letting a dividing cell be instantly replaced by two identical progeny cells that inherit the spatial position and phenotypic state of the parent cell. Conversely, a cell undergoing death is instantly removed from the population. Focusing on a scenario wherein the cell population undergoes density-dependent inhibition of growth, the probabilities of cell proliferation and death are assumed here to also depend on the cell density. Hence, a focal cell in the phenotypic state $y_j$ at spatial position $x_i$ at time $t_k$ will proliferate with probability $P^P(t_k, x_i, y_j,\rho^k_i)$, die with probability $P^D(t_k, x_i, y_j,\rho^k_i)$, or remain quiescent (i.e neither proliferate nor die) with probability $P^Q(t_k, x_i, y_j,\rho^k_i)= 1-P^P(t_k, x_i, y_j,\rho^k_i)-P^D(t_k, x_i, y_j,\rho^k_i)$. In particular, modelling the rate of cell proliferation and the rate of cell death through the functions $r^P$ and $r^D$, respectively, we use the following definitions
{\small
\begin{equation}
\label{def:PPPD}
P^P(t_k, x_i, y_j,\rho^k_i) := \tau \, r^P(t_k, x_i,y_j,\rho^k_i) \ , \quad P^D(t_k, x_i, y_j,\rho^k_i) := \tau \, r^D(t_k, x_i,y_j,\rho^k_i) \ ,
\end{equation}
}
under the additional assumption that the time-step $\tau$ is sufficiently small so that $P^P(t_k,x_i,y_j,\rho^k_i) + P^D(t_k,x_i,y_j,\rho^k_i) \leq 1$ for all $(t_k,x_i,y_j,\rho^k_i) \in \mathbb{R}^+ \times \mathbb{R} \times \mathbb{R}  \times \mathbb{R}^+$.

\subsubsection{Formal derivation of the corresponding PS-PDE model}
\label{sec:derfromIBmodel}
Assuming the time-step, $\tau$, the space-step, $h_x$, and the phenotype-step, $h_y$, to be sufficiently small, we formally take $t := t_k$, $x := x_i$, $y := y_j$, and thus $t_{k+1} := t + \tau$, $x_{i \pm 1} := x \pm h_x$, $y_{j \pm 1} := y \pm h_y$, and $n(t_k,x_i,y_j):=n(t,x,y)$. Furthermore, recalling that $\rho(t_k,x_i)$ is defined via~\eqref{eq:nNIB}, we also formally take $\rho(t_k,x_i) := \rho(t,x)$ with $\displaystyle{\rho(t,x) := \int_{\mathbb{R}} n(t,x,y) \ {\rm d}y}$. Then, when the dynamics of single cells are governed by the rules summarised above, under the simplifying assumption that the events underlying spatial movement, phenotypic changes, and cell proliferation and death are independent, the principle of mass balance formally gives the following equation
{\small
\begin{eqnarray*}
n(t + \tau,x,y) &=& \Big(2 \, P^P(t,x,y,\rho(t,x)) + P^Q(t,x,y,\rho(t,x)) \Big) \times
\\
&& \; \times \Big \{\underbrace{p_y^S \, P^S_x(t,x,y) \, n(t,x,y)}_{\substack{\mbox{\scriptsize{cells remaining at $(x,y)$}}}} +
\\
&& \;\;\;\,\, + \underbrace{p_y^L \, P^S_x(t,x,y+h_y) \, n(t,x,y+h_y) +  p_y^R \, P^S_x(t,x,y-h_y) \, n(t,x,y-h_y)}_{\substack{\mbox{\scriptsize{cells switching from $(x,y \pm h_y)$ to $(x,y)$}}}} + 
\\
&& \;\;\;\,\, + \underbrace{p_y^S \, \Big[P^L_x(t,x+h_x,y) \, n(t,x+h_x,y) + P^R_x(t,x-h_x,y) \, n(t,x-h_x,y)\Big]}_{\substack{\mbox{\scriptsize{cells switching from $(x\pm h_x,y)$ to $(x,y)$}}}} + 
\\
&& \;\;\;\,\, + \underbrace{p_y^L \, P^R_x(t,x-h_x,y+h_y) \, n(t,x-h_x,y+h_y)}_{\substack{\mbox{\scriptsize{cells switching from $(x- h_x,y+h_y)$ to $(x,y)$}}}} +
\\
&& \;\;\;\,\, + \underbrace{p_y^L \, P^L_x(t,x+h_x,y+h_y) \, n(t,x+h_x,y+h_y)}_{\substack{\mbox{\scriptsize{cells switching from $(x+h_x,y+h_y)$ to $(x,y)$}}}} +
\\
&& \;\;\;\,\, + \underbrace{p_y^R \, P^R_x(t,x-h_x,y-h_y) \, n(t,x-h_x,y-h_y)}_{\substack{\mbox{\scriptsize{cells switching from $(x- h_x,y-h_y)$ to $(x,y)$}}}} + 
\\
&& \;\;\;\,\, + \underbrace{p_y^R \, P^L_x(t,x+h_x,y-h_y) \, n(t,x+h_x,y-h_y)}_{\substack{\mbox{\scriptsize{cells switching from $(x + h_x,y-h_y)$ to $(x,y)$}}}} \Big \} \ .
\end{eqnarray*}
}
When the probabilities of cell proliferation and death are defined via~\eqref{def:PPPD}, using the fact that $P^Q(t, x, y,\rho(t,x))= 1-P^P(t, x, y,\rho(t,x))-P^D(t, x, y,\rho(t,x))$, we have 
{\small
$$
2 \, P^P(t, x, y,\rho(t,x)) + P^Q(t, x, y,\rho(t,x)) =  1 + \tau \, \left(r^P(t,x,y,\rho(t,x)) - r^D(t,x,y,\rho(t,x)) \right) \ .
$$
}
Hence, introducing the following definition of the net cell proliferation rate (i.e. the difference between the rate of cell proliferation and the rate of cell death)
$$
R(t,x,y,\rho) := r^P(t,x,y,\rho) - r^D(t,x,y,\rho) \ ,
$$
rewriting $\rho(t,x)$ as $\rho$ for brevity, and using the fact that $P^S_x(t,x,y) = 1 - P^R_x(t,x,y) - P^L_x(t,x,y)$ and $p^S_y = 1 - p^R_y - p^L_y$, the above equation can be rewritten as
{\small
\begin{eqnarray*}
n(t + \tau,x,y) &=& \Big(1 + \tau \, R(t,x,y,\rho) \Big) \times 
\\
&& \;\; \times \Big \{\left(1 - p^R_y - p^L_y \right) \Big(1 -  P^R_x(t,x,y) - P^L_x(t,x,y)\Big) n(t,x,y) +
\\
&& \;\;\;\,\, + p_y^L \, \Big(1 -  P^R_x(t,x,y+h_y) - P^L_x(t,x,y+h_y)\Big) \, n(t,x,y+h_y) +
\\
&& \;\;\;\,\, + p_y^R \, \Big(1 -  P^R_x(t,x,y-h_y) - P^L_x(t,x,y-h_y)\Big) \, n(t,x,y-h_y) + 
\\
&& \;\;\;\,\, + \left(1 - p^R_y - p^L_y \right) \, \Big[P^R_x(t,x-h_x,y) \, n(t,x-h_x,y) +
\\
&& \;\;\;\,\, \phantom{+ \left(1 - p^R_y - p^L_y \right) \, \Big[} + P^L_x(t,x+h_x,y) \, n(t,x+h_x,y) \Big] + 
\\
&& \;\;\;\,\, + p_y^L \, \Big[P^R_x(t,x-h_x,y+h_y) \, n(t,x-h_x,y+h_y) +
\\
&& \;\;\;\,\, \phantom{+ p_y^L \, \Big[}+ P^L_x(t,x+h_x,y+h_y) \, n(t,x+h_x,y+h_y) \Big]
\\
&& \;\;\;\,\, + p_y^R \, \Big[P^R_x(t,x-h_x,y-h_y) \, n(t,x-h_x,y-h_y) + 
\\
&& \;\;\;\,\, \phantom{+ p_y^R \, \Big[}+ P^L_x(t,x+h_x,y-h_y) \, n(t,x+h_x,y-h_y) \Big] \Big \}.
\end{eqnarray*}
}
Provided that the functions $P^R_x(t,x,y)$, $P^L_x(t,x,y)$, and $n(t,x,y)$ are sufficiently regular, one can substitute the first-order Taylor polynomial for $n(t+\tau,x,y)$ and the second-order Taylor polynomials for the other terms 
into the above equation. Dropping $(t,x,y)$ for brevity, after some simple (but tedious) calculations we are left with the following terms (arranged in increasing order of $\tau$, $h_x$, and $h_y$)
{\small
\begin{eqnarray*}
\cancel{n} +\tau\,\partial_t n + O(\tau^2) &=& 1\times \Big\{ \, \cancel{n} + h_x\, \partial_x\Big[ n\left(P^L_x-P^R_x\right)\Big]  + h_y\, (p^L_y-p^R_y) \, \partial_y n 
+ 
\\
&& \hspace{20pt}+ \dfrac{h_x^2}{2} \,\partial^2_{xx}\Big[\left(P^R_x+P^L_x\right)n\Big] + 
\dfrac{h_y^2}{2} \, \left(p^R_y+p^L_y\right) \, \partial^2_{yy}n + 
\\
&& \hspace{20pt}+
h_xh_y\, \left(p^R_y-p^L_y\right)\partial^2_{xy}\Big[\left(P^R_x-P^L_x\right)n\Big] + O(h_x^2h_y) + O(h_xh_y^2)\Big\} \, +
\\
&& + \,\tau\,R(t,x,y,\rho) \times \Big\{\, n \, + O(h_x) + O(h_y) \, \Big\} + \ h.o.t. \, ,
\end{eqnarray*}
}
where the terms in $\{\cdot\}$ come from the same brackets in the previous equation, and higher order terms in $\tau$, $h_x$, and $h_y$ have been grouped into $h.o.t. \,$. Dividing through by $\tau$ and rearranging terms, we find the following equation for the cell phenotype density $n$
{\small
\begin{eqnarray}
\label{eq:deriv1}
\partial_t n &=& R(t,x,y,\rho) \, n + \nonumber
\\
&& + \partial_x \left[n \, \left(\dfrac{h_x}{\tau} \left(P^L_x - P^R_x\right) + \dfrac{h_x^2}{2 \tau} \partial_x \left(P^L_x + P^R_x\right) \right) + \dfrac{h_x^2}{2 \tau} \left(P^L_x + P^R_x\right) \partial_x n \right] + \nonumber
\\
&& + \ \dfrac{h_y}{\tau} \left(p^L_y - p^R_y\right) \ \partial_y n + \dfrac{h_y^2}{2 \tau} \ \left(p^L_y + p^R_y\right) \ \partial^2_{yy} n + \nonumber
\\
&& + \ \dfrac{h_x h_y}{\tau} \  \left(p^L_y - p^R_y\right) \partial^2_{xy} \left[\left(P^L_x - P^R_x\right)   n  \right] + \ h.o.t. \, , 
\end{eqnarray}
}
where higher order terms in $\tau$, $h_x$, and $h_y$ have again been grouped into $h.o.t. \,$. If the functions $P^L_x$ and $P^R_x$ and the parameters $p^L_y$ and $p^R_y$ are also such that the following relations hold as $h_x \to 0$ and $h_y \to 0$ 
$$
P^L_x(t,x,y) - P^R_x(t,x,y) = \dfrac{h_x}{2} \Psi_x(t,x,y) + \mathcal{O}(h_x^2) \ , \quad  \Psi_x : \mathbb{R}^+ \times \mathbb{R} \times \mathbb{R} \to \mathbb{R} \ ,
$$
$$
P^L_x(t,x,y) + P^R_x(t,x,y) = \Phi_x(t,x,y) + \mathcal{O}(h_x) \ , \quad  \Phi_x : \mathbb{R}^+ \times \mathbb{R} \times \mathbb{R} \to \mathbb{R}^+ \ ,
$$
and
$$
p^L_y - p^R_y = \frac{h_y}{2}\psi_y \,  + \mathcal{O}(h_y^2) \ , \qquad p^L_y + p^R_y = \phi_y + \mathcal{O}(h_y) \ , \qquad \psi_y \in \mathbb{R} \ ,  \phi_y \in \mathbb{R}^+ \ , 
$$
where $\Psi_x$ and $\Phi_x$ are sufficiently regular functions, then letting $\tau \to 0$, $h_x \to 0$, and $h_y \to 0$ in such a way that 
$$
\dfrac{h_x^2}{2 \tau} \to \alpha_x \in \mathbb{R}^+_* \ , \qquad \dfrac{h_y^2}{2 \tau} \to \alpha_y \in \mathbb{R}^+_* \ ,
$$
from equation~\eqref{eq:deriv1} we formally obtain the following equation
\begin{eqnarray*}
\partial_t n &=& R(t,x,y,\rho) \, n +  \partial_x \left[n \, \alpha_x \, \left(\Psi_x(t,x,y) + \partial_x \Phi_x(t,x,y) \right)\right] +
\\
&& 
+ \partial_{x} \left[\alpha_x \, \Phi_x(t,x,y) \partial_x n\right] +  \alpha_y \, \psi_y \partial_y n + \alpha_y \, \phi_y \, \partial^2_{yy} n \ . 
\end{eqnarray*}
Introducing the definitions
$$
A_x(t,x,y) := -\alpha_x \, \left(\Psi_x(t,x,y) + \partial_x \Phi_x(t,x,y) \right) \ , \quad D_x(t,x,y):= \alpha_x \, \Phi_x(t,x,y) \ ,
$$
and
$$
\bar{A} := - \alpha_y \ \psi_y \ , \quad \bar{D} := \alpha_y \, \phi_y \ ,
$$
and recalling the definition of $\rho(t,x)$ given above, from the latter equation we find the following PS-PDE model
\begin{equation*}
\begin{cases}
\displaystyle{\partial_t n + \partial_x \left[A_x(t,x,y) \, n - D_x(t,x,y) \, \partial_x n \right] + \partial_y \left[\bar{A} \, n - \bar{D} \, \partial_y n \right] = n \, R(t,x,y,\rho) \ ,}\\[5pt]
\displaystyle{\rho(t,x) := \int_{\mathbb{R}} n(t,x,y) \, {\rm d}y \ ,}
\end{cases}
\end{equation*}
which is of the form of the PS-PDE model~\eqref{general}, posed on a one-dimensional unbounded physical and phenotype domain, with 
$$
A_y(t,x,y) \equiv \bar{A} \ , \quad D_y(t,x,y)\equiv \bar{D} \ , \quad F(t,x,y) := n(t,x,y) \, R(t,x,y,\rho(t,x)) \ .
$$
Choosing $\psi_y=0$, so that $\bar{A}=0$, under different definitions of the functions $\Psi_x(t,x,y)$ and $\Phi_x(t,x,y)$, we can derive many of the models stated previously in Section \ref{pdetopspide}, including: 
\begin{itemize}
    \item the case where $$
\Psi_x(t,x,y) \equiv 0 \ , \qquad \Phi_x(t,x,y) \equiv \Phi_x(y) := \hat D(y) \ ,
$$
from which, defining $D(y) := \alpha_x \hat D(y)$ and taking also $R(t,x,y,\rho) \equiv R(y,\rho)$, we obtain a PS-PDE of form~\eqref{structuredFKPP};
\item[]
\item the case where $$
\Psi_x(t,x,y) :=  \hat \mu(y) \, \partial_x P(t,x), \qquad \Phi_x(t,x,y) \equiv 0 \ , \qquad P(t,x):= \Pi[\rho](t,x) \ ,
$$
from which, defining $\mu(y) := \alpha_x \hat \mu(y)$ and taking $R(t,x,y,\rho) \equiv R(y,\Pi[\rho]) \equiv R(y,P)$, we obtain a PS-PDE of form~\eqref{structuredPM};
\item[]
\item the case where $$
\Psi_x(t,x,y) := - \, \hat \chi(y) \, \partial_x S(t,x) \ , \qquad \Phi_x(t,x,y) \equiv \hat D \ ,
$$
from which, defining $\chi(y) := \alpha_x \hat \chi(y)$ and $D := \alpha_x \hat D$, and taking also $R(t,x,y,\rho) \equiv R(y,\rho,S(t,x))$, we obtain a PS-PDE of form~\eqref{structuredKS}.
\end{itemize}

As a concluding remark we note that an excellent agreement can be found between the outputs of the ABM and the solutions of the corresponding PS-PDE model, if the functions and parameters of the ABM are such that the conditions underlying the formal derivation carried out here are satisfied; an example is provided by the comparison between the plots in Figure~\ref{pressurebasedfigure}(a) and Figure~\ref{pressurebasedfigure}(b). This holds, in particular, when sufficiently large cell populations are considered. Then, demographic stochasticity (which cannot be captured by PS-PDE models) does not play a dominant role in the ABM's cell dynamics.

\subsection{Tools and techniques to analyse PS-PDE models}
\label{sec:tools:analysis}
\newcommand{\e}{\varepsilon}
In the last decade, considerable attention has been directed towards analysing the behaviour of the solutions to PS-PDE models of type~\eqref{general}, with particular emphasis on travelling waves and concentration phenomena. Referring back to Figures~\ref{pressurebasedfigure} and~\ref{chemotaxisfigure} -- which display the results of numerical simulations of appropriately rescaled versions of the PS-PDEs~\eqref{structuredPM} and~\eqref{structuredKS} -- 
we can see phenotypic structuring across travelling waves as a result of the interplay between these two characteristic features: the cell densities $\rho(t,x)$ corresponding to solutions to these models behave like travelling waves (Figure~\ref{pressurebasedfigure}(b), right panel and Figure~\ref{chemotaxisfigure}, middle panels); within these waves, a concentration phenomenon occurs, with the phenotype density $n(t,x,y)$ being concentrated into a different phenotypic state $y$ at different spatial positions $x$ across the wave (Figure~\ref{pressurebasedfigure}(b), left panel and Figure~\ref{chemotaxisfigure}, left panels).

The obtained results, as well as the tools and techniques through which these results have been realised, build upon the mathematical literature on related PDE and non-local PDE models. Therefore, in order to make this section (to an extent) self-contained, before reaching PS-PDE models (in Section~\ref{sec:analysis:TWCPPSPIDE}), we first recall the essentials of travelling waves in PDE models of spatial spread dynamics (in Section~\ref{sec:analysis:TWPDE}) and concentration phenomena in non-local PDE models of evolutionary dynamics (in Section~\ref{sec:analysis:CPNPDE}), taking the PDE~\eqref{FKPP} and the non-local PDE~\eqref{nonlocalFKPP} as prototypical examples, respectively. 

\subsubsection{Travelling waves in PDE models of spatial spread dynamics}
\label{sec:analysis:TWPDE}
When exploring spatial spread dynamics through PDE models of type~\eqref{dar} with $x \in \mathbb{R}$, one is often led to study travelling waves, which are solutions that propagate without a change in shape and at a constant speed $c \in \mathbb{R}$ (see Figure \ref{FKPPfigure}), and are thus of the form
\begin{equation}\label{eq:TWdef}
\rho(t,x) \equiv \rho(z)\footnote{Note that we use the same notation $\rho$ for the function of $(t,x)$ and of $z$ even though, formally, they are not the same function. This is to avoid introducing an extra notation at this stage.} \ , \quad z = x - c \, t \ , \quad z \in (-\infty,\infty) \  .
\end{equation}
In particular, focusing on travelling waves which propagate to the right, we take $c \in \mathbb{R}^+_*$. Investigating whether the model admits travelling wave solutions then amounts to proving the existence of pairs $(\rho(z),c)$, such that $\rho : (-\infty,\infty) \to \mathbb{R}^+_*$ satisfies the ODE obtained by substituting~\eqref{eq:TWdef} into the PDE for $\rho(t,x)$, subject to suitable boundary conditions at $z=-\infty$ and/or $z=\infty$, for some $c \in \mathbb{R}^+_*$.  

For instance, substituting~\eqref{eq:TWdef} into the Fisher-KPP model~\eqref{FKPP}-\eqref{def:RFKPP} and rearranging terms gives the following ODE for $\rho(z)$ 
\begin{equation}
\label{eq:ODETW}
D \, \frac{{\rm d}^2 \rho}{{\rm d}z^2}+  c \frac{{\rm d}\rho}{{\rm d}z} + r \ \rho\left(1-\dfrac{\rho}{k}\right)  = 0 \ , \quad  z \in \mathbb{R} \ .
\end{equation}
Moreover, since the Fisher-KPP model~\eqref{FKPP}-\eqref{def:RFKPP} admits the homogeneous steady-state solutions $\rho \equiv k$ and $\rho \equiv 0$, which can be proven to be linearly asymptotically stable and unstable, respectively, it is natural to complement the ODE~\eqref{eq:ODETW} with the boundary conditions 
\begin{equation}
\label{eq:ODETWBCs}
\rho(-\infty)=k \ , \quad \rho(\infty)=0 \ ,
\end{equation}
so that the stable homogeneous steady-state ``invades'' the unstable one. The boundary conditions~\eqref{eq:ODETWBCs} then correspond to considering a scenario in which the population spreads across the uncolonised surrounding space. A classical result is that there exist monotonically decreasing solutions of the problem~\eqref{eq:ODETW}-\eqref{eq:ODETWBCs}, provided that $c \geq c^* := 2 \ \sqrt{r D}$, where the threshold value of the speed $c^*$ is called the minimal wave speed~\citep{murray2003mathematical,perthame2015parabolic}.

When a travelling-wave solution $\rho(z)$ exists, one may then be interested in investigating whether the solution $\rho(t,x)$ to the Cauchy problem defined by the PDE model with $x \in \mathbb{R}$ and complemented with an appropriate initial condition will converge to a solution  qualitatively similar to the travelling-wave solution for large $t$ and $x$. A technique that can be used to address this aspect consists of employing the following space and time scaling
\begin{equation}\label{an:scaling:fisher}
(t,x)\rightarrow \left( \frac{t}{\e}, \frac{x}{\e} \right) \,,
\end{equation}
and investigating the behaviour of $\displaystyle{\rho_\e(t,x) \equiv \rho\left(t/\e, x/\e \right)}$ in the limit $\e\to0$~\citep{evans1989pde,freidlin1986geometric}. For instance, choosing $D=r=k=1$, under the scaling~\eqref{an:scaling:fisher} the Fisher-KPP model~\eqref{FKPP}-\eqref{def:RFKPP} with $x \in \mathbb{R}$ reads as
\begin{equation}\label{FisherKPP:rescaled}
\e \partial_t \rho_\e =\e^2  \partial^2_{xx} \rho_\e + \rho_\e (1 - \rho_\e)\,, \quad x \in \mathbb{R}\,.
\end{equation}
As $\e \to 0$, the limit, $\rho(z)$, of the travelling wave solution, $\rho_{\e}(z)$, for the rescaled Fisher-KPP equation~\eqref{FisherKPP:rescaled} satisfies the relation $\rho(z) \ (1-\rho(z))=0$. Hence, it attains only the values $1$ and $0$. Moreover, the corresponding minimal wave speed is $c^* = 2$. In accordance with this, it is possible to prove that, when $\e\to0$, the solution of the Cauchy problem defined by complementing the rescaled Fisher-KPP equation~\eqref{FisherKPP:rescaled} with the initial condition 
\begin{equation}\label{FisherKPP:rescaledIC}
\rho_{\e}(0,x) = \rho^0(x) :=  
\begin{cases}
1, \quad \text{if } x<0
\\
0, \quad \text{if } x \geq 0
\end{cases}
\end{equation}
converges (in some appropriate sense) to $\rho^0(x-c^* t)$, with $c^*=2$~\citep{evans1989pde,freidlin1986geometric}. A corroboration of this result is shown in Figure \ref{FKPPfigure}, where numerical solutions of~\eqref{FisherKPP:rescaled}-\eqref{FisherKPP:rescaledIC} are seen to become increasingly step-like as $\e\rightarrow 0$ and with a numerically calculated wave speed $v \approxeq 2$. As shown by~\cite{barles1990wavefront,fleming1986pde,evans1989pde}, a useful tool to prove this and related asymptotic results is the real phase WKB ansatz borrowed from geometric optics, which is also referred to as the Hopf-Cole transformation:
\begin{equation}\label{wkb:fisher}
\rho_\e(t,x) = \exp \left( \frac{u_\e(t,x)}{\e} \right)  \,.
\end{equation}
A key observation underlying the change of variable~\eqref{wkb:fisher} is that if $u_\e(t,x)$ converges (uniformly, locally in time) to $u(t,x)$ as $\e\to0$, then $\rho(t,x)=0$ and $\rho(t,x)=1$ correspond to $u(t,x)<0$ and $u(t,x)=0$, respectively. Hence, the limiting behaviour of $\rho_\e(t,x)$ can be characterised by studying the limiting behaviour of $u_\e(t,x)$. The advantage of this approach lies in the fact that upon substituting the ansatz~\eqref{wkb:fisher} into the rescaled (local) PDE~\eqref{FisherKPP:rescaled}, one finds that $u_\e(t,x)$ satisfies a type of Hamilton-Jacobi equation for which there is a range of mathematical tools, stemming from the work of Lions and coworkers in the 1980s on viscosity solutions of Hamilton-Jacobi equations~\citep{crandall1983viscosity,crandall1984some,lions1982generalized}. This in turn facilitates studies into the behaviour of $\rho_\e(t,x)$ as $\e \to 0$. 

\subsubsection{Concentration phenomena in non-local PDE models of evolutionary dynamics}
\label{sec:analysis:CPNPDE}

A commonly made assumption within theoretical studies into the evolution of phenotype-structured populations is that phenotype density functions are Gaussians~\citep{rice2004evolutionary}. Hence, it is natural to consider non-local PDE models for evolutionary dynamics in phenotype-structured populations of form~\eqref{nonlocalFKPP} with $y \in \mathbb{R}$ subject to initial conditions of the form 
\begin{equation}
\label{eq:ICNLFKPP}
n(0,y) = n^0(y) := \frac{\rho^0}{\sqrt{2 \pi \, \sigma_0^2}} \, \exp\left[-\frac{\left(y - \bar{y}^0 \right)^2}{2 \, \sigma_0^2}\right] \,, \quad \rho^0\,, \sigma_0 \in \mathbb{R}^+_{*} \,, \;\; \bar{y}^0 \in \mathbb{R} \,.
\end{equation}
Here, the parameters $\rho^0$ and $\bar{y}^0$ model the initial size of the population and the mean or prevailing phenotype in the population at time $t=0$, respectively. The parameter $\sigma_0^2$ is the related variance, which provides a measure of intra-population phenotypic heterogeneity. It is possible to prove (see, for instance, ~\citealt{almeida2019evolution,ardavseva2020evolutionary,chisholm2016evolutionary}) that, when subject to the initial condition~\eqref{eq:ICNLFKPP}, the non-local PDE model~\eqref{nonlocalFKPP}-\eqref{def:pEx1} with $y \in \mathbb{R}$ admits solutions of the Gaussian form
\begin{equation}
\label{eq:SOLNLFKPP}
n(t,y) = \frac{\rho(t)}{\sqrt{2 \pi \, \sigma^2(t)}} \, \exp\left[-\frac{\left(y - \bar{y}(t) \right)^2}{2 \, \sigma^2(t)}\right] \, ,
\end{equation}
where the size of the cell population, $\rho(t)$, the mean or prevailing phenotype, $\bar{y}(t)$, and the inverse of the related variance, $v(t) = 1/\sigma^2(t)$, satisfy the Cauchy problem
\begin{equation}
\label{eq:NLFKPPodes}
\left\{
\begin{array}{ll}
\displaystyle{\frac{{\rm d} v}{{\rm d}t} = 2 \left(1 - \overline{D} v^2\right)} \ ,
\\\\
\displaystyle{\frac{{\rm d} \bar{y}}{{\rm d}t} = \frac{2}{v} \left(\varphi - \bar{y}\right)} \ , 
\\\\
\displaystyle{\frac{{\rm d} \rho}{{\rm d}t} = \left\{\left[\gamma - \frac{1}{v} - \left(\bar{y} - \varphi \right)^2\right] - \kappa \ \rho \right\}\rho} \ ,
\\\\
v(0) = 1/\sigma^2_0 \ , \quad \bar{y}(0) = \bar{y}^0 \ , \quad \rho(0) = \rho^0 \ .
\end{array}
\right.
\end{equation} 
An exhaustive quantitative characterisation of the dynamics of the phenotype density $n(t,y)$ can then be obtained by analysing the behaviour of the components of the solution to the Cauchy problem~\eqref{eq:NLFKPPodes}. 

Furthermore, when studying evolutionary dynamics in phenotype-structured populations, one is often interested in predicting the phenotypic composition of the population over long timescales in scenarios wherein: (i) proliferation and death play a leading role in phenotypic evolution, since they drive adaptation by natural selection; (ii) phenotypic changes are rare, and their role is thus limited to generating the substrate for natural selection to act upon~\citep{burger2000mathematical,perthame2006transport}. In the framework of mathematical models of form~\eqref{nonlocalFKPP}, this can be done by using the parameter scaling 
\begin{equation}
\label{an:scaling:overlineD}
\overline{D} := \e^2
\end{equation}
along with the time scaling $t \to t/\e$, and then investigating the behaviour of $n_{\e}(t,y) \equiv n(t/\e,y)$ in the asymptotic regime $\e \to 0$~\citep{diekmann2005dynamics,perthame2006transport,perthame2008dirac}. Under this scaling, the non-local PDE~\eqref{nonlocalFKPP} with $y \in \mathbb{R}$ reads as 
\begin{equation}\label{nonlocal-FisherKPP:rescaled}
\begin{cases}
\displaystyle{\e\partial_t  n_\e =\e^2  \partial^2_{yy}  n_\e + n_\e \, R(y,\rho_\e) \,, \quad y \in \mathbb{R}\,,}\\[5pt]
\displaystyle{\rho_\e(t) := \int_{\mathbb{R}} n_\e(t,y)\,{\rm d}y\,.}
\end{cases}
\end{equation}
In the case when there is little phenotypic variability in the population at $t=0$, which is a case often considered in adaptive dynamics~\citep{diekmann2004beginner}, one can also assume that intra-population phenotypic heterogeneity is initially small and then use the additional parameter scaling 
\begin{equation}
\label{ass:sigma0eps}
\sigma^2_0:=\e
\end{equation}
in~\eqref{eq:ICNLFKPP}, which gives the initial condition
\begin{equation}
\label{eq:ICNLFKPPrev}
n_{\e}(0,y) = n^0_{\e}(y) := \frac{\rho^0}{\sqrt{2 \pi \e}} \, \exp\left[-\dfrac{\left(y - \bar{y}^0 \right)^2}{2 \e}\right] \,, \quad \rho^0 \in \mathbb{R}^+_{*} \,, \;\; \bar{y}^0 \in \mathbb{R} \,.
\end{equation}
From~\eqref{eq:SOLNLFKPP} and~\eqref{eq:NLFKPPodes}, one sees that, when subject to the initial condition~\eqref{eq:ICNLFKPPrev}, the rescaled non-local PDE obtained by complementing~\eqref{nonlocal-FisherKPP:rescaled} with~\eqref{def:REx1},\eqref{def:pEx1} admits the solution
\begin{equation}
\label{eq:SOLNLFKPPreveps}
n_{\e}(t,y) = \frac{\rho_{\e}(t)}{\sqrt{2 \pi \e}} \, \exp\left[-\frac{\left(y - \bar{y}_{\e}(t) \right)^2}{2 \e} \right] \, , 
\end{equation}
where
 \begin{equation}
\label{eq:NLFKPPodesreveps}
\left\{
\begin{array}{ll}
\displaystyle{\frac{{\rm d} \bar{y}_{\e}}{{\rm d}t} = 2  \left(\varphi - \bar{y}_{\e}\right)}, 
\\\\
\displaystyle{\e \, \frac{{\rm d} \rho_{\e}}{{\rm d}t} = \left\{\left[\gamma - \e - \left(\bar{y}_{\e} - \varphi \right)^2\right] - \kappa \ \rho_{\e} \right\}\rho_{\e}},
\\\\
\bar{y}_\e(0) = \bar{y}^0, \quad \rho_{\e}(0) = \rho^0 \ .
\end{array}
\right.
\end{equation} 

Under the additional assumption that $\bar{y}_0$ is such that $(\bar{y}_0 -\varphi)^2<\gamma$,  letting $\e \to 0$ in~\eqref{eq:SOLNLFKPPreveps} and~\eqref{eq:NLFKPPodesreveps} one finds 
 \begin{equation}
 \label{eq:concphen}
n_{\e}(t,y) \xrightharpoonup[\e  \rightarrow 0]{}\rho(t) \delta_{\bar{y}(t)}(y) \quad \text{(weakly in measures)} \, , 
\end{equation}
where $\bar{y}(t)$ satisfies the Cauchy problem
 \begin{equation}
\label{eq:NLFKPPbaryeq}
\left\{
\begin{array}{ll}
\displaystyle{\frac{{\rm d} \bar{y}}{{\rm d}t} = 2  \left(\varphi - \bar{y}\right)}\,, 
\\\\
\bar{y}(0) = \bar{y}^0\,,
\end{array}
\right.
\end{equation}
and $\rho(t)>0$ is given as a function of $\bar{y}(t)$ by the relation
 \begin{equation}
\label{eq:NLFKPPrhoeq}
\rho(t) = \dfrac{\gamma - \left(\bar{y}(t) - \varphi \right)^2}{\kappa} \;\; \text{for a.e.} \;\; t \in (0,\infty) \, .
\end{equation}
This is also confirmed by the results of numerical simulations of~\eqref{nonlocal-FisherKPP:rescaled}, subject to~\eqref{eq:ICNLFKPPrev} and complemented with~\eqref{def:REx1},\eqref{def:pEx1}, which are displayed in Figure~\ref{fig:gaussiansol}. Here, numerical solutions are seen to be of the Gaussian form~\eqref{eq:SOLNLFKPPreveps}, which becomes concentrated as a weighted infinitely sharp Gaussian (i.e. a weighted Dirac mass) in the asymptotic regime $\e\rightarrow 0$.

The concentration phenomenon expressed by the asymptotic result~\eqref{eq:concphen} -- together with the assumptions placed for the initial condition, fitness function, and parameter regime for the rescaling -- provides a mathematical formalisation for the following idea: when phenotypic changes are rare and the fitness function has a single maximum point (i.e. there is only one phenotype with maximal fitness -- the `fittest' phenotype) then, if a phenotype-structured population is monomorphic (shows only one trait) at  $t=0$, the population will remain monomorphic at all times $t>0$. 

Moreover, as proposed by~\cite{diekmann2005dynamics}, the concentration point (i.e. the centre of the Dirac mass) $\bar{y}(t)$ can be biologically interpreted as the trait that is shown by the population at time $t$. Hence, the ODE~\eqref{eq:NLFKPPbaryeq}$_1$ can be regarded as a type of canonical equation of adaptive dynamics: an ODE that describes how the prevailing trait in the population changes over time~\citep{metz1986dynamics}. From the Cauchy problem~\eqref{eq:NLFKPPbaryeq} it is easy to see that $\bar{y}(t) \to \varphi$ as $t \to \infty$. Recalling that the parameter $\varphi$ models the fittest phenotype (see the fitness function defined via~\eqref{def:REx1},\eqref{def:pEx1}), the latter convergence result translates into mathematical terms the concept of  ``survival of the fittest'' originating from evolutionary theory~\citep{spencer2020principles}.

Notably, the solution to the rescaled non-local PDE~\eqref{nonlocal-FisherKPP:rescaled} also exhibits concentration phenomena for a broader class of fitness functions and initial conditions, provided they have essentially the same structural properties of the fitness function defined via~\eqref{def:REx1},\eqref{def:pEx1} and the initial condition~\eqref{eq:ICNLFKPPrev} -- in other words, functions $R(y,\rho)$ that are strictly concave in $y$ and strictly monotonically decreasing in $\rho$, and initial conditions $n^0_{\e}(y)$ that converge (in the weak sense of measures) to weighted Dirac masses as $\e \to 0$. 

A robust method for studying such concentration phenomena was originally proposed by~\cite{diekmann2005dynamics,perthame2006transport}, and then developed and extended  by~\cite{barles439concentrations,barles2009concentration,chisholm2016effects,lam2017dirac,lorz2017dirac,lorz2011dirac,mirrahimi2012singular,perthame2008dirac}. This built on the observation that, given that the rescaled non-local PDE~\eqref{nonlocal-FisherKPP:rescaled}$_1$ is of close form to the rescaled Fisher-KPP equation~\eqref{FisherKPP:rescaled}, with the exception of how the reaction term is defined, it is natural to make a change of variable of the same type as~\eqref{wkb:fisher}, that is, use the WKB ansatz
\begin{equation}\label{wkb:nonlocalfisher}
n_\e(0,y) = \exp \left( \frac{u_\e(t,y)}{\e} \right)  \,.
\end{equation}
We refer the interested reader to~\cite{perthame2014some} for an effective summary of this method, which, in brief, consists of deriving a priori $L^{\infty}$- and $BV$-estimates for $\rho_{\e}(t)$ and then analysing the Hamilton-Jacobi equation for $u_\e(t,y)$ obtained by substituting~\eqref{wkb:nonlocalfisher} into~\eqref{nonlocal-FisherKPP:rescaled}$_1$. Through this method, it is possible to prove that the asymptotic result~\eqref{eq:concphen} still holds, under appropriate assumptions on the fitness function and the initial condition, including that
$$
n_\e(t,y) = n^0_{\e}(y) := \exp \left(\frac{u^0_\e(y)}{\e} \right) \, ,
$$
where $u^0_\e(y)$ is a strictly concave function such that
$$
n^0_{\e}(y) \xrightharpoonup[\e  \rightarrow 0]{}\rho^0 \delta_{\bar{y}^0}(y) \quad \text{(weakly in measures)} \, , \quad \rho_0 \in \mathbb{R}^+_{*}\,, \;\; \bar{y}_0 \in \mathbb{R} \,.
$$ 
Under this more general scenario, the concentration point $\bar{y}(t)$ in~\eqref{eq:concphen} becomes the solution to the Cauchy problem
 \begin{equation}
\label{eq:NLFKPPbaryeqgen}
\left\{
\begin{array}{ll}
\displaystyle{\frac{{\rm d} \bar{y}}{{\rm d}t} = - \dfrac{\partial_y R(\bar{y},\rho)}{\partial^2_{yy} u(t,\bar{y})}} \,, 
\\\\
\bar{y}(0) = \bar{y}^0\,.
\end{array}
\right.
\end{equation}
Here $u(t,y)$ -- which is the limit of $u_{\e}(t,y)$ as $\e \to 0$ -- is a viscosity solution (in the sense introduced by~\cite{barles439concentrations,barles2009concentration,perthame2008dirac}) of the constrained Hamilton-Jacobi equation
\begin{equation}\label{eq:HJgen}
\begin{cases}
\displaystyle{\partial_t u(t,y) = (\partial_{y}  u(t,y))^2 + R(y,\rho(t))  \,, \quad y \in \mathbb{R}\,  \,,}
\\
\displaystyle{\max_{y\in \mathbb{R}} u(t,y) = u(t,\bar{y}(t))=0\,,}
\end{cases}
\end{equation}
subject to the initial condition $u(0,y) = u^0(y)$ (i.e. the limit of $u^0_{\e}(y)$ as $\e \to 0$), and it is such that $\partial^2_{yy} u(t,\bar{y}(t))<0$. Furthermore, under the assumption that the function $R(y,\rho)$ is strictly monotonically decreasing in $\rho$, and thus invertible, the weight $\rho(t)>0$ in~\eqref{eq:concphen} is given as a function of $\bar{y}(t)$ by the relation
\begin{equation}\label{eq:FKPP:R:SSgen}
    R(\bar{y}(t),\rho(t))=0 \;\; \text{for a.e.} \;\; t \in (0,\infty) \, ,
\end{equation} 
and it can also be regarded as a Lagrange multiplier associated with the constraint~\eqref{eq:HJgen}$_2$. Note that considerations analogous to those we made on the ODE~\eqref{eq:NLFKPPbaryeq}$_1$ apply to the ODE~\eqref{eq:NLFKPPbaryeqgen}$_1$ as well. Specifically, the ODE~\eqref{eq:NLFKPPbaryeqgen}$_1$ can be interpreted as a type of canonical equation of adaptive dynamics and it is such that, when $R(y,\rho)$ is a strictly concave function of $y$ that attains its maximum at $\varphi \in \mathbb{R}$ (i.e. $\varphi$ is the fittest phenotype), if $\bar{y}(t)$ converges to some $\bar{y}^{\infty} \in \mathbb{R}$ as $t \to \infty$ then $\bar{y}^{\infty}=\varphi$.    

As final remarks, the aforementioned results on Gaussian solutions and concentration phenomena extend to the case where the fitness function is also a periodic function of $t$, which corresponds to the situation of populations exposed to periodically fluctuating environments, as demonstrated, for instance, by~\cite{ardavseva2020evolutionary,lorenzi2015dissecting} and~\cite{figueroa2018long,mirrahimi2015time}. Furthermore, concentration phenomena have been investigated also in non-local PDE models of evolutionary dynamics of forms similar to~\eqref{nonlocalFKPP} but where the diffusion term is replaced by an advection term modelling cell differentiation, with an advection velocity that is a function of the phenotype structure~\citep{guilberteau2023long}.

\subsubsection{Travelling waves and concentration phenomena in PS-PDE models of spatial and evolutionary dynamics}
\label{sec:analysis:TWCPPSPIDE}

\paragraph{Travelling waves}
In the context of PS-PDE models of type~\eqref{general}, travelling waves are solutions of the form
\begin{equation}
\label{eq:TWdefevo}
n(t,x,y) \equiv n(z,y) \ , \quad z = x - c \, t \ , \quad z \in (-\infty,\infty) \  ,
\end{equation}
which generalises the form~\eqref{eq:TWdef} to the case where a phenotypic structure is incorporated into the model. Building on the formal results presented by~\cite{bouin2012invasion}, particular attention has been given to travelling-wave solutions of the PS-PDE model~\eqref{eq:CT}-\eqref{def:RDCT} with ${\bm x} \equiv x \in \mathbb{R}$ and ${\bm y} \equiv y \in (0,Y)$, where $Y \in \mathbb{R}^+_*$, and subject to zero-flux boundary conditions at $y=0$ and $y=Y$.

Summarising, the existence of solutions to the boundary value problem defined by complementing the PDE obtained by substituting~\eqref{eq:TWdefevo} into the PS-PDE model~\eqref{eq:CT}-\eqref{def:RDCT} with appropriate boundary conditions at $z=-\infty$ and $z=\infty$ was proven by~\cite{bouin2014travelling}, who also characterised the corresponding minimal wave speed. The convergence of the solution to the PS-PDE model~\eqref{eq:CT}-\eqref{def:RDCT} to travelling waves, for large $t$ and $x$ and under appropriate assumptions on the initial condition, was subsequently studied by~\cite{turanova2015model}. The first step of the study carried out by~\cite{turanova2015model} employs the space and time scaling~\eqref{an:scaling:fisher}, which leads to the following rescaled PS-PDE model
\begin{equation}\label{eq:rescCaTo}
\begin{cases}
\displaystyle{\e\partial_t  n_\e = \e^2 \ D(y) \ \partial^2_{xx} n_{\e} + \overline{D}  \partial^2_{yy}  n_\e + n_\e \, R(\rho_{\e}) \,, \quad x \in \mathbb{R} \ , \;  y \in (0,Y)\,,}\\[5pt]
\displaystyle{\rho_\e(t,x) := \int_{0}^Y n_\e(t,x,y)\,{\rm d}y} \ ,
\end{cases}
\end{equation} 
complemented with~\eqref{def:RDCT}. Then, one studies the limiting behaviour as $\e \to 0$ of the solution to the Hamilton-Jacobi equation which is obtained by substituting the WKB ansatz 
\begin{equation}\label{wkb:xy}
n_\e(t,x,y) = \exp \left( \frac{u_\e(t,x,y)}{\e} \right)  
\end{equation}
into~\eqref{eq:rescCaTo}$_1$. Note that~\eqref{wkb:xy} is a natural extension of~\eqref{wkb:fisher} to the case where the solution to the model equation depends both on the spatial variable $x$ and the phenotypic variable $y$. Convergence of the solution to the PS-PDE model~\eqref{eq:CT}-\eqref{def:RDCT} to travelling waves as $t \to \infty$, again under appropriate assumptions on the initial condition, was also studied by~\cite{berestycki2015existence,bouin2017super}, who established a detailed characterisation of the dynamics of the maximum point of the phenotype density function $n$ along the phenotypic dimension $y$ (i.e. the dynamics of the prevailing phenotype) at the edge of the wave front as well. The emergence of accelerating fronts when $Y=\infty$, which, for the choice of $D(y)$ given by~\eqref{def:RDCT} (i.e. $D(y):=y$), corresponds to the case of unbounded motility, was also investigated  by~\cite{berestycki2015existence,bouin2012invasion,bouin2017super}. 

\paragraph{Concentration phenomena}
A natural generalisation of Gaussian initial conditions of form~\eqref{eq:ICNLFKPP} to the case when spatial dynamics are also taken into account is provided by initial conditions of the form 
\begin{equation}
\label{eq:ICNLFKPPhetenv}
n(0,x,y) = n^0(x,y) := \frac{\rho^0(x)}{\sqrt{2 \pi \, \sigma_0^2(x)}} \, \exp\left[-\frac{\left(y - \bar{y}^0(x) \right)^2}{2 \, \sigma_0^2(x)}\right] 
\end{equation}
with
$$
\rho^0 : \mathcal{X} \to \mathbb{R}^+_{*} \ , \quad \bar{y}^0 : \mathcal{X} \to \mathbb{R} \ , \quad \sigma_0 : \mathcal{X} \to \mathbb{R}^+_{*} \ ,
$$
where the functions $\rho^0(x)$, $\bar{y}^0(x)$, and $\sigma^2_0(x)$ model, respectively, the cell density, the mean or prevailing phenotype of cells at position $x$, and the related variance at time $t=0$.

\cite{villa2021evolutionary} proved that the PS-PDE model~\eqref{nonlocal-FisherKPPhetenv}-\eqref{def:pEx1hetenv} with ${\bm x} \equiv x \in \mathcal{X} \subset \mathbb{R}$, ${\bm y} \equiv y \in \mathbb{R}$, $D(y)\equiv0$ (i.e. cell movement is neglected), and ${\bm S}(t,{\bm x}) \equiv {\bm \Sigma}(x)$, where ${\bm \Sigma}(x)$ is given (i.e. the spatially heterogeneous environment in which the population is embedded does not evolve in time), and subject to initial condition~\eqref{eq:ICNLFKPPhetenv},
admits solutions of the Gaussian form
\begin{equation}
\label{eq:SOLNLFKPPhetenv}
n(t,x,y) = \frac{\rho(t,x)}{\sqrt{2 \pi \, \sigma^2(t,x)}} \, \exp\left[-\frac{\left(y - \bar{y}(t,x) \right)^2}{2 \, \sigma^2(t,x)}\right] \, .
\end{equation}
Here the cell density, $\rho(t,x)$, the mean or prevailing phenotype at position $x$, $\bar{y}(t,x)$, and the inverse of the related variance, $v(t,x) = 1/\sigma^2(t,x)$, satisfy a Cauchy problem analogous to~\eqref{eq:NLFKPPodes}, that is,
$$
\left\{
\begin{array}{ll}
\displaystyle{\partial_t v = 2 \left(1 - \overline{D} v^2\right)} \ ,
\\\\
\displaystyle{\partial_t \bar{y} = \frac{2}{v} \left(f(x) - \bar{y}\right)} \ , 
\\\\
\displaystyle{\partial_t \rho = \left\{\left[g(x) - \frac{1}{v} - \left(\bar{y} - f(x) \right)^2\right] - \kappa \ \rho \right\}\rho} \ ,
\\\\
v(0,x) = 1/\sigma^2_0(x) \ , \quad \bar{y}(0,x) = \bar{y}^0(x) \ , \quad \rho(0,x) = \rho^0(x) \ , 
\end{array}
\right.
\qquad 
x \in \mathcal{X} \, ,
$$
where $f(x) \equiv f[{\bm \Sigma}](x)$ and $g(x) \equiv g[{\bm \Sigma}](x)$. As a result, under the space and time scaling~\eqref{an:scaling:fisher}, the parameter scaling~\eqref{an:scaling:overlineD}, and the scaling $\sigma^2_0(x)\equiv \e$ (which generalises the parameter scaling~\eqref{ass:sigma0eps} to the case where spatial dynamics are also taken into account), given 
appropriate assumptions on the functions $g(x)$, $f(x)$, and $\bar{y}^0(x)$, it is natural to expect concentration phenomena along the phenotypic dimension $y$ to emerge at each spatial position $x$. 

In fact, concentration phenomena have also been studied for rescaled PS-PDE models of form
\begin{equation}\label{nonlocal-FisherKPPhetenv:rescaled}
\begin{cases}
\displaystyle{\e\partial_t  n_\e = \e^2 D \partial^2_{xx}  n_\e + \e^2  \partial^2_{yy}  n_\e + n_\e \, R(y,\rho_\e,S_{\e}) \,, \quad x \in \mathcal{X} \ , \; y \in \mathcal{Y}\,,}\\[5pt]
\displaystyle{\rho_\e(t,x) := \int_{\mathcal{Y}} n_\e(t,x,y)\,{\rm d}y\,,}
\end{cases}
\end{equation}
which can be obtained from PS-PDE models like~\eqref{nonlocal-FisherKPPhetenv}, when ${\bm x} \equiv x \in \mathcal{X}$, ${\bm y} \equiv y \in \mathcal{Y}$, $D({\bm y}) \equiv D$, and ${\bm S}(t,{\bm x}) \equiv S(t,x)$, by employing the space and time scaling~\eqref{an:scaling:fisher} alongside the parameter scaling~\eqref{an:scaling:overlineD}. 

To provide greater detail, using the WKB ansatz~\eqref{wkb:xy} -- a natural choice in view of the similarities between the rescaled PS-PDE~\eqref{nonlocal-FisherKPPhetenv:rescaled}$_1$ and the rescaled non-local PDE~\eqref{nonlocal-FisherKPP:rescaled}$_1$ -- and making appropriate assumptions on the initial condition, $n_\e(0,x,y)$, the fitness function, $R$, and the governing equation for $S_{\e}$, asymptotic results can be obtained that express concentration phenomena analogous to~\eqref{eq:concphen}, i.e. 
\begin{equation}
 \label{eq:concphenspace}
n_{\e}(t,x,y) \xrightharpoonup[\e  \rightarrow 0]{}\rho(t,x) \delta_{\bar{y}(t,x)}(y) \quad \text{(weakly in measures)} \, .
\end{equation}
These were proven by~\cite{jabin2023selection}, building on the results presented by~\cite{mirrahimi2015asymptotic} for the case where $D=0$ and $\mathcal{Y} \equiv \mathbb{R}$. In addition, when $\mathcal{X} \equiv \mathbb{R}$, assuming
$$
n^0_{\e}(y) \xrightharpoonup[\e  \rightarrow 0]{}\rho^0(x) \delta_{\bar{y}^0(x)}(y) \quad \text{(weakly in measures)} \, , \quad \rho^0 : \mathbb{R} \to \mathbb{R}^+_{*} \ , \quad \bar{y}_0 : \mathbb{R} \to \mathbb{R} \ ,
$$
one can formally show, as performed by~\cite{villa2021modeling} taking $D>0$, that the concentration point $\bar{y}(t,x)$ in~\eqref{eq:concphenspace} formally satisfies the following Cauchy problem
\begin{equation}
\label{eq:NLFKPPbaryeqgenspacegen}
\left\{
\begin{array}{ll}
\displaystyle{\partial_t \bar{y} = - \dfrac{\partial_y R(\bar{y},\rho,S)}{\partial^2_{yy} u(t,x,\bar{y})}} \,, 
\\\\
\bar{y}(0,x) = \bar{y}^0(x)\,, 
\end{array}
\right.
\qquad 
x\in \mathbb{R} \, .
\end{equation}
In~\eqref{eq:NLFKPPbaryeqgenspacegen}$_1$, the function $S$ is the limit of $S_{\e}$ as $\e \to 0$ and, analogously to~\eqref{eq:NLFKPPbaryeqgen}, the function $u(t,x,y)$ , which is the limit of $u_{\e}(t,x,y)$ as $\e \to 0$, satisfies a Hamilton-Jacobi equation subject to the constraint
$$
\displaystyle{\max_{y\in \mathbb{R}} u(t,x,y) = u(t,x,\bar{y}(t,x))=0} \ , 
\quad (t,x) \in (0,\infty) \times \mathbb{R} \ ,
$$
and it is such that $\partial^2_{yy} u(t,x,\bar{y}(t,x))<0$. Moreover, if the function $R(y,\rho,S)$ is strictly monotonically decreasing in $\rho$, the weight $\rho(t,x)>0$ in~\eqref{eq:concphenspace} is given as a function of both $\bar{y}(t,x)$ and $S(t,x)$ by the relation
\begin{equation}\label{eq:FKPP:R:SSspacegen}
    R(\bar{y}(t,x),\rho(t,x),S(t,x))=0 \ , \;\; \text{for a.e.} \;\; t \in (0,\infty) \ .
\end{equation} 

From a biological point of view, the asymptotic result~\eqref{eq:concphenspace} extends the asymptotic result~\eqref{eq:concphen} to the case of phenotype-structured populations in spatially heterogeneous environments, whereby the concentration point $\bar{y}(t,x)$ represents the trait that is shown by the population at time $t$ and position $x$ (i.e. the locally prevailing phenotype). Furthermore, for each $x$, the ODE~\eqref{eq:NLFKPPbaryeqgenspacegen}$_1$ can be regarded as a generalised canonical equation of adaptive dynamics. When $R(y,\rho,S)$ is a strictly concave function of $y$, with maximum point given by the function $f[S](t,x)$ (i.e. $f[S](t,x)$ is the fittest phenotype at position $x$ and time $t$ depending on the environmental conditions determined by $S(t,x)$), if $S(t,x)$ and $\bar{y}(t,x)$ converge, respectively, to some $S^{\infty}(x)$ and $\bar{y}^{\infty}(x)$ for every $x$ as $t \to \infty$ then $\bar{y}^{\infty}(x)=f[S^{\infty}](x)$. This formalises in mathematical terms the idea that the fittest phenotype, which is determined by the local environmental conditions, is ultimately selected at each spatial position.

As a concluding remark, we note that concentration phenomena of type~\eqref{eq:concphenspace} have also been investigated in PS-PDE models of form~\eqref{nonlocal-FisherKPPhetenv} with $D({\bm y}) \equiv D$, see~\citep{bouin2015hamilton}, and in more general PS-PDE models of related forms, see~\citep{hao2019concentration}.

\paragraph{Concentration phenomena across travelling waves}
More recently, concentration phenomena have been investigated across travelling waves in PS-PDE models for the spatial spread and evolutionary dynamics of cell populations under the pressure-based form~\eqref{structuredPM} and the taxis-based form~\eqref{structuredKS}, with ${\bm x} \equiv x \in \mathbb{R}$ and ${\bm y} \equiv y \in (0,Y)$ where $Y \in \mathbb{R}^+_*$. Applying the space and time scaling~\eqref{an:scaling:fisher} and the parameter scaling~\eqref{an:scaling:overlineD}, alongside the scaling $\mu(y) \to \dfrac{\mu(y)}{\e}$ for~\eqref{structuredPM} and the scaling $\chi(y) \to \dfrac{\chi(y)}{\e}$ for~\eqref{structuredKS}, rescaled models of the type of~\eqref{structuredKS} are given by
\begin{equation}\label{PSPIDES:rescaled}
\begin{cases}
\displaystyle{\e\partial_t  n_\e + \e \ \partial_x \left[n_\e \chi(y) \partial_{x} S_{\e}  - \e D \partial_{x}  n_\e\right] =  \e^2  \partial^2_{yy}  n_\e + n_\e \, R(y,\rho_\e,S_{\e}) \,, \; x \in \mathbb{R} \ , \; y \in (0,Y)\,,}\\[5pt]
\displaystyle{\rho_\e(t,x) := \int_{0}^Y n_\e(t,x,y)\,{\rm d}y} \ ,
\end{cases}
\end{equation} 
complemented with an equation governing the dynamics of $S_{\e}(t,x)$, and rescaled models of the type of~\eqref{structuredPM} are given by
\begin{equation}\label{PSPIDEP:rescaled}
\begin{cases}
\displaystyle{\e\partial_t  n_\e - \e \ \partial_x \left[n_\e \mu(y) \partial_{x} P_{\e}  \right] =  \e^2  \partial^2_{yy}  n_\e + n_\e \, R(y,P_{\e}) \,, \quad x \in \mathbb{R} \ , \; y \in (0,Y)\,,}\\[5pt]
\displaystyle{P_{\e}(t,x) := \Pi[\rho_\e](t,x) \ , \quad \rho_\e(t,x) := \int_{0}^Y n_\e(t,x,y)\,{\rm d}y}\,.
\end{cases}
\end{equation} 
Both rescaled PS-PDEs are subject to zero-flux boundary conditions at $y=0$ and $y=Y$.

In particular, under appropriate assumptions on the model functions, including that the function $R$ is strictly monotonically decreasing in $\rho$, employing the ansatz~\eqref{wkb:xy}, it was shown that rescaled PS-PDEs of form~\eqref{PSPIDES:rescaled}, see~\citep{lorenzi2022trade,lorenzi2023derivation}, and rescaled PS-PDEs of form~\eqref{PSPIDEP:rescaled}, see~\citep{lorenzi2022invasion,macfarlane2022individual}, formally admit travelling-wave solutions $n_{\e}(t,x,y) \equiv n_{\e}(z,y)$, with $c \in \mathbb{R}^+_*$, such that
\begin{equation}
 \label{eq:concphenspaceTW}
n_{\e}(z,y) \approx \rho(z) \delta_{\bar{y}(z)}(y)\,, \quad \text{as } \e \to 0 \,.
\end{equation} 
These correspond to biological scenarios wherein the cell population is monomorphic at each position along the wave, and the concentration point $\bar{y}(z)$ and the weight $\rho(z)$ represent, respectively, the trait that is expressed by the cells and the cell density at position $z$. For the rescaled PS-PDE model~\eqref{PSPIDES:rescaled},  $\bar{y}(z)$ formally satisfies the ODE
\begin{equation}
\label{eq:TWSbary}
\left(c - \chi(\bar{y}) \frac{{\rm d}S}{{\rm d}z} \right) \frac{{\rm d}\bar{y}}{{\rm d}z} = \frac{\partial_{y} R(\bar{y},\rho,S)}{\partial^2_{yy} u(z,\bar{y})}, \quad z \in {\rm Supp}\left(\rho \right) \ ,
\end{equation}
with ${\rm Supp}(\rho):= \{z \in \mathbb{R} : \rho(z)>0\}$. Moreover, $\rho(z)>0$ is formally given as a function of $\bar{y}(z)$ and $S(z)$ by the relation
\begin{equation}
\label{eq:TWSrho}
    R(\bar{y}(z),\rho(z),S(z))=0 \ , \quad z \in {\rm Supp}(\rho) \ .
\end{equation} 
The ODE~\eqref{eq:TWSbary} and the relation~\eqref{eq:TWSrho} are coupled with an ODE for $S(z)$, which is the travelling-wave solution of the equation for $S_{\e}(t,x)$ as $\e \to 0$. Analogously, for the rescaled PS-PDE model~\eqref{PSPIDEP:rescaled}, $\bar{y}(z)$ formally satisfies the ODE
\begin{equation}
\label{eq:TWPbary}
\left(c + \mu(\bar{y}) \frac{{\rm d}P}{{\rm d}z} \right) \frac{{\rm d}\bar{y}}{{\rm d}z} = \frac{\partial_{y} R(\bar{y},P)}{\partial^2_{yy} u(z,\bar{y})}, \quad z \in {\rm Supp}\left(P \right) \ ,
\end{equation}
and $P(z)>0$ and $\rho(z)>0$ are formally given as functions of $\bar{y}(z)$ by the relations
\begin{equation}
\label{eq:TWPPrho}
    R(\bar{y}(z),P(z))=0 \ , \quad \rho(z) = \Pi^{-1}[P](z) \ ,\quad z \in {\rm Supp}(P) \ ,
\end{equation} 
where $\Pi^{-1}$ denotes the inverse of the function $\Pi$. The existence of this is ensured by the following assumptions on $\Pi(\rho)$
\begin{equation}
\Pi(0)=0 \ ,\quad \frac{\mathrm{d}}{\mathrm{d}\rho} \Pi(\rho) > 0 \; \text{ for }\; \rho \in \mathbb{R}^+_* \, ,
\label{pressure_cond}
\end{equation}
which are common assumptions to make~\citep{ambrosi2002closure}. The function $u(z,y)$ in~\eqref{eq:TWSbary} (or~\eqref{eq:TWPbary}) formally satisfies a Hamilton-Jacobi equation subject to the constraint
$$
\displaystyle{\max_{y\in [0,Y]} u(z,y) = u(z,\bar{y}(z))=0} \ , \quad z \in {\rm Supp}(\rho) \left(\text{or }  z \in {\rm Supp}(P)\right)\ ,
$$
and is such that $\partial^2_{yy} u(z,\bar{y}(z))<0$. 

Moreover, under definition~\eqref{def:REx1hetenv} taking $\kappa=\dfrac{1}{\rho_M}$ with $\rho_M \in \mathbb{R}^+_*$, i.e. defining 
$$
R(y,S,\rho) \equiv R(y,\rho) := r(y) - \dfrac{\rho}{\rho_M} \ ,
$$
and under definition~\eqref{def:REP} taking $\kappa=\dfrac{1}{P_M}$ with $P_M \in \mathbb{R}^+_*$, i.e. defining
$$
R(y,P) := r(y) -  \dfrac{P}{P_M} \ ,
$$
the relations~\eqref{eq:TWSrho} and~\eqref{eq:TWPPrho} reduce, respectively, to
\begin{equation}
\label{eq:TWSrhored}
    \rho(z) = \rho_M \, r(\bar{y}(z)) \ , \quad z \in {\rm Supp}(\rho) \ ,
\end{equation} 
and
\begin{equation}
\label{eq:TWPPred}
    P(z) = P_M \, r(\bar{y}(z)) \ , \quad \rho(z) = \Pi^{-1}[P](z) \ ,\quad z \in {\rm Supp}(P) \ .
\end{equation} 
Furthermore, one can set monotonicity assumptions on the functions $r(y)$, $\chi(y)$, and $\mu(y)$ that are relevant to biological scenarios in which the inherent energetic cost attached to cellular activities leads to proliferation-migration trade-offs: $y$ close to $0$ corresponds to high proliferation and low migration abilities, while $y$ close to $Y$ corresponds to low proliferation and high migration abilities. Thus, we assume the functions $\chi$ and $\mu$ to be monotonically increasing on $(0,Y)$ and such that $0<\chi(0)<\chi(Y)<\infty$ and $0<\mu(0)<\mu(Y)<\infty$, while the function $r$ is monotonically decreasing on $(0,Y)$ and such that $r(0)=1$ and $r(Y)=0$. Considering the bulk of the population to be at $z=-\infty$, it is then natural to complement the ODEs~\eqref{eq:TWSbary} and~\eqref{eq:TWPbary} with the boundary condition 
\begin{equation}
\label{eq:TWbaryBC}
\bar{y}(-\infty) = 0 \ ,
\end{equation} 
so that the relations~\eqref{eq:TWSrhored} and~\eqref{eq:TWPPred} give, respectively,
\begin{equation}
\label{eq:TWrhoBC}
\rho(-\infty) = \rho_M
\end{equation} 
and
\begin{equation}
\label{eq:TWPBC}
P(-\infty) = P_M \ , \quad \rho(-\infty) = \Pi^{-1}(P_M) =: \rho_M \ .  
\end{equation} 
Note that assumptions~\eqref{pressure_cond} ensure that if $0 \leq P \leq P_M$ then $\Pi^{-1}(P_M)$ is the maximum value of $\rho(z) = \Pi^{-1}[P](z)$. Under the boundary condition~\eqref{eq:TWbaryBC}, through direct calculations on \eqref{eq:TWSbary},\eqref{eq:TWSrho} and~\eqref{eq:TWPbary},\eqref{eq:TWPPrho}, it is possible to  show that, in both cases, there is a minimal wave speed $c^* \in \mathbb{R}^+_*$ such that if $c>c^*$ then the concentration point $\bar{y}(z)$ and the weight $\rho(z)$ in~\eqref{eq:concphenspaceTW} formally satisfy the following relations 
$$
\bar{y}(z) = 0 \;\; \wedge \;\; \bar{y}'(z) > 0 \ , \;\; z \in (-\infty, \ell) \;\; \wedge \;\;  \bar{y}(\ell) = Y
$$
and
$$
\rho(-\infty) = \rho_M \; \wedge \; \rho'(z) < 0 \ , \;\; z \in (-\infty, \ell) \; \wedge \;  \rho(z) = 0 \ , \;\; z \in [\ell, \infty) \ ,
$$
with $\ell \in \mathbb{R} \cup \{\infty\}$. Under the aforementioned monotonicity assumptions on the functions $r(y)$, $\chi(y)$, and $\mu(y)$, these results provide a mathematical formalisation of the idea that cells with a more migratory and less proliferative phenotype (i.e. in phenotypic states corresponding to $y$ close to $Y$) are concentrated towards the front of the travelling wave, whereas cells with a less migratory and more proliferative phenotype (i.e. in phenotypic states corresponding to $y$ close to $0$) make up the population bulk in the rear of the wave. Numerical simulations of the rescaled pressure-based PS-PDE model~\eqref{PSPIDEP:rescaled} and the rescaled taxis-based PS-PDE~model~\eqref{PSPIDES:rescaled} show this within-wave phenotype structuring (see Figure~\ref{pressurebasedfigure}(b), left panel and Figure~\ref{chemotaxisfigure}(a),  left panel), where we also note the verification of the minimal wave speed (see Figure~\ref{pressurebasedfigure}(b), inset of the right panel and Figure~\ref{chemotaxisfigure}(a), right panel).
    
      \subsection{Tools and techniques to simulate PS-PDE models}
    \label{sec:tools:numerics}
    When simulating \psipdes of form~\eqref{general}, numerical challenges may arise due to the potential composite shapes and sharp features of the solution, or the stiff\footnote{A stiff PDE generally contains terms that can lead to rapid variations in the solution, to the extent that  methods for numerical integration in time would generally require extremely small time-steps to ensure stability.  This concept naturally extends to PDE systems, where dynamics modelled by different equations may occur over different timescales.} nature of the problems.  We offer a brief step-by-step guide into the techniques that can be used to overcome these challenges, when employing numerical schemes that rely on the method of lines (MOL)~\citep{hundsdorfer2003numerical}. We illustrate through the following one-dimensional, in each of physical and phenotype space, PS-PDE model
     \begin{equation}\label{eq:numerics}
\begin{cases}
\displaystyle{\partial_t n = \partial_{x}  \left[D(y) \partial_x n - A(y,\rho,S) n \right] + \bar{D} \partial^2_{yy} n +  n R(y,\rho,S)\,, \; x \in {\cal {X}} \ , \; y \in {\cal {Y}},}\\[5pt]
\displaystyle{\rho(t,x) := \int_{{\cal {Y}}} n(t,x,y) \, {\rm d}{ y} \ .}
\end{cases}
\end{equation}
We note that from this PS-PDE model we can obtain: the PS-PDE model~\eqref{structuredFKPP} by taking $A\equiv0$ and $R(y,\rho,S)\equiv R(y,\rho)$; the PS-PDE model~\eqref{structuredPM} with $\Pi(\rho):=\rho$ by taking $D\equiv0$, $A(y,\rho,S) \equiv A(y,\rho) :=-\mu(y)\partial_x\rho$, and again $R(y,\rho,S)\equiv R(y,\rho)$; and the PS-PDE model~\eqref{structuredKS} by taking $D(y)\equiv D$ and $A(y,\rho,S) \equiv A(y,S):=\chi(y)\partial_x S$. Python code to solve the PS-PDE model~\eqref{structuredFKPP} using the scheme described below is available in Open Access (see `Code Availability' at the end of the manuscript).
    
The MOL has been the most popular method for simulating PS-PDE models, and involves a discretisation of the \psipde in both phenotype and physical space to obtain a high-dimensional system of ODEs. The solution of this system is tracked over a computational mesh and integrated in time through an appropriate method.

    \paragraph{Discretisation in phenotype and physical space}
    Naturally, numerical solutions based on the MOL require bounded domains\footnote{If one needs to simulate a PS-PDE on an unbounded domain, some alternative approach -- such as a pseudospectral method, see e.g. \citep{fornberg1998practical} --  would need to be invoked.}, i.e. $\mathcal{X}\subset\mathbb{R}$ and $\mathcal{Y}\subset\mathbb{R}$ in~\eqref{eq:numerics}.  Consider a uniform discretisation of the domain  $\mathcal{X}\times\mathcal{Y}$, comprising of $N_x\times N_y$ grid cells each of area $\Delta x \times \Delta y$ and with cell centres $\{x_i\}_{i=1,...,N_x}\times\{y_j\}_{j=1,...,N_y}$, where $x_i = i\Delta x - \frac{1}{2}\Delta x$ ($i=1,...,N_x$) and $y_j = j\Delta y - \frac{1}{2}\Delta y$ ($j=1,...,N_y$). We then consider a numerical approximation for the average of $n$, $S$, and $\rho$ across each cell, and let
    \begin{equation*}\label{num:discrete}
    n(t,x_i,y_j)\approx n_{i,j}(t)\,, \quad \rho(t,x_i)\approx \rho_i(t):= \Delta y \sum_{j=1}^{N_y}n_{i,j}(t)\,, \quad S(t,x_i)\approx S_i(t)\,,
    \end{equation*}
    where the definition of $\rho_i$ relies on a discretisation of the integral~\eqref{eq:numerics}$_2$ using a middle Riemann sum.  Then, each $n_{i,j}$ satisfies an ODE in the form
    \begin{equation}\label{num:ode}
    \frac{\rm d}{{\rm d}t} n_{i,j}(t) = M_{i,j}(t) + C_{i,j}(t) + n_{i,j}(t) R_{i,j}(t)\,,
    \end{equation}
    where $M_{i,j}(t)$ is defined by the approximation of the term that models movement through physical space in~\eqref{eq:numerics}, $C_{i,j}(t)$ is the approximation of the term that models phenotypic changes, and $R_{i,j}(t):= R(y_j,\rho_i(t),S_i(t))$.  
    Adopting the finite volume method~\citep{eymard2000finite}, the term $M_{i,j}$ is defined via
     \begin{equation*}\label{num:M}
    M_{i,j}(t) := \, \frac{1}{\Delta x}\left[ F^*_{i+\frac{1}{2},j}(t) -  F^*_{i-\frac{1}{2},j}(t) \right]  , \qquad i=2,...,N_{x-1},\;\; j=1,...,N_y,
    \end{equation*}
     where $F^*_{i+\frac{1}{2},j}(t)$ and $F^*_{i-\frac{1}{2},j}(t)$ represent the numerical approximations of the flux through physical space along the cell boundary $\left\{x_i+\frac{1}{2}\Delta x\right\}\times \left(y_i-\frac{1}{2}\Delta y, y_i+\frac{1}{2}\Delta y\right)$ and along the cell boundary $\left\{x_i-\frac{1}{2}\Delta x\right\}\times \left(y_i-\frac{1}{2}\Delta y, y_i+\frac{1}{2}\Delta y\right)$, respectively. Applying a first-order central finite difference approximation for the first-order derivative and a first-order upwind approximation for the advection term~\citep{david2022asymptotic,lorenzi2022invasion}, these are given by
     \begin{align}
         &F^*_{i+\frac{1}{2},j}(t) := D(y_j)\frac{n_{i+1,j}-n_{i,j}}{\Delta x} - \left( A_{i+\frac{1}{2},j}(t)\right)_+ n_{i,j}(t)  + \left(A_{i+\frac{1}{2},j}(t)\right)_- n_{i+1,j}\,,\label{num:upwind1}\\[5pt]
         &F^*_{i-\frac{1}{2},j}(t) := D(y_j)\frac{n_{i,j}-n_{i-1,j}}{\Delta x} - \left( A_{i-\frac{1}{2},j}(t)\right)_+ n_{i-1,j}(t)  + \left(A_{i-\frac{1}{2},j}(t)\right)_- n_{i,j}\,,\label{num:upwind2}
     \end{align}
   where we used the notation $(\cdot )_+=\max (0,\cdot)$ and $(\cdot )_-=\min (0,\cdot)$. The terms $A_{i+\frac{1}{2},j}(t)$ and $A_{i-\frac{1}{2},j}(t)$ represent the approximation of $A(y,\rho,S)$ along the cell boundary of interest, and their definition will be problem-dependent. For instance,  in~\eqref{structuredPM} these may be given by
   \begin{equation*}
       A_{i+\frac{1}{2},j}:= - \mu(y_j) \frac{\rho_{i+1}-\rho_{i}}{\Delta x}\quad \text{and} \quad A_{i-\frac{1}{2},j}:= - \mu(y_j) \frac{\rho_{i}-\rho_{i-1}}{\Delta x}\,,
   \end{equation*}
   while in~\eqref{structuredKS} they may be given by 
   \begin{equation*}
       A_{i+\frac{1}{2},j}:=  \chi(y_j) \frac{S_{i+1}-S_{i}}{\Delta x} \quad \text{and} \quad A_{i-\frac{1}{2},j}:=  \chi(y_j) \frac{S_{i}-S_{i-1}}{\Delta x}\,.
   \end{equation*}
Following similar steps, the approximation $C_{i,j}(t)$ of the linear diffusion term in phenotype space is given by the three-point stencil scheme
    \begin{equation*}\label{num:C}
    C_{i,j}(t) :=\bar{D} \,\frac{n_{i,j-1}(t) - 2 n_{i,j} (t)+ n_{i,j+1}(t)}{(\Delta y)^2}, \quad i=1,...,N_x,\;\; j=2,...,N_{y-1},
    \end{equation*}
    which is analogous to the typical second-order central finite difference approximation of second-order derivatives.  The definition of $M_{i,j}(t)$ and $C_{i,j}(t)$ at the remaining boundary cells will depend on the chosen boundary conditions, and can follow from first principles of the finite volume method or may be imposed through the addition of `ghost points'.
    
    The use of a first-order upwind scheme in~\eqref{num:upwind1} and~\eqref{num:upwind2} is particularly helpful as a means to avoid the emergence of spurious oscillations, especially given that solutions can display large gradients or even discontinuities -- e.g. in the absence of linear diffusion. These features may arise as concentration phenomena can occur in phenotype space (i.e. $n$ becomes concentrated as a sharp Gaussian along the $y$-dimension, see {Figure~\ref{fig:gaussiansol}), while travelling fronts or patterns with sharp interfaces can emerge in physical space~\citep{lorenzi2022invasion,lorenzi2022trade,lorenzi2023derivation}, see for instance Figure~\ref{chemotaxisfigure}. In some cases a first-order upwind scheme may not suffice and alternative discretisations -- e.g. using flux limiters, such as the MUSCL scheme~\citep{van1979towards} -- may be more appropriate. 

     Another strategy to circumvent the issues induced by the lack of regularity that arise if the solution concentrates into a Dirac mass (e.g. as when introducing the scalings~\eqref{an:scaling:fisher} and~\eqref{an:scaling:overlineD} and considering the asymptotic regime $\e \to 0$) is to exploit the WKB ansatz~\eqref{wkb:xy}, or a related ansatz, and solve the equation for $u_{\e}$ instead, so as to define an asymptotic preserving scheme~\citep{almeida2022asymptotic,calvez2023concentration}. 

 Given the potential for the solution to exhibit sharp features, an important step in mesh selection is to check the solution accuracy via testing different levels of mesh refinement. Alternatively, one may opt for a non-uniform mesh that concentrates grid points in areas of large gradients~\citep{kolbe2014numerical,kolbe2022adaptive}. Of course, mesh refinement and coarsening algorithms inevitably lead to increased computational costs.

        \paragraph{Time integration}
        Following the discretisation in phenotype and physical space,  one needs to solve the system of $N_x\times N_y$ ODEs~\eqref{num:ode} by integrating in time.  Consider a uniform discretisation of the time interval $[0,T]$ with time-step of size $\Delta t$, i.e. $t_k = k\Delta t$ ($k=0,...,N_t$).  Introducing the notation $n^k_{i,j}:=n_{i,j}(t_k)$ -- and similarly for the other time-dependent variables -- a first-order forward difference scheme yields 
        \begin{equation*}
        n^{k+1}_{i,j} = n^{k}_{i,j} + \Delta t \left[ M^k_{i,j} + C^k_{i,j} + n^k_{i,j}R^k_{i,j}\right]\,.
        \end{equation*}
        This is the simplest explicit scheme that one may adopt, being first-order in time; higher-order schemes may be more appropriate if the problem requires higher accuracy, e.g. when investigating features as those considered in~\citep{lorenzi2022trade}.  Nonetheless, stiffer problems may require an excessively small value for $\Delta t$ to ensure stability in an explicit scheme, e.g. due to a severe CFL (Courant–Friedrichs–Lewy) constraint~\citep{hundsdorfer2003numerical}. If so, an implicit scheme may be preferable and, as an example, a first-order backward difference for the time derivative yields
        \begin{equation*}
        n^{k+1}_{i,j} = n^{k}_{i,j} + \Delta t \left[ M^{k+1}_{i,j} + C^{k+1}_{i,j} + n^{k+1}_{i,j}R^{k+1}_{i,j}\right]\,.
        \end{equation*}
        This is a simple example of a fully implicit scheme which is unconditionally stable, but may remain computationally expensive -- especially in higher dimensions -- due to the requirement of inverting large matrices at each iteration.  
        
        One method to lower the computational cost associated with implicit solvers is through  a time-splitting scheme: treating stiff parts of the problem implicitly, and solving the remaining parts explicitly~\citep{macfarlane2022individual,lorenzi2022invasion}. In the case of simple forms of spatial movement (e.g. linear diffusion) one may, for instance, treat the reaction term implicitly and the conservative part of the equation -- which can often occur on a slower timescale -- explicitly, e.g. see~\citep{lorenzi2015dissecting}. However, the precise choice is inevitably problem-dependent and more complex forms of cellular motion may also require an implicit treatment~\citep{lorenzi2022invasion}. 
    
    The use of implicit solvers is also particularly useful for preserving non-negativity of solutions: crucial, given the usual biological meaning of $n$ and -- generally -- the structural properties of the \psipde. Explicit solvers may indeed fail at this, for example if the negative part of $R$ becomes particularly large. Implicit-explicit approaches can also be adopted to circumvent this issue, for instance by splitting the reaction term to treat the negative part of $nR$ implicitly and the positive part explicitly~\citep{lorenzi2015dissecting,lorz2011dirac}. Time-splitting schemes relying on exploiting the WKB ansatz and solving the equation for $u$ instead, in the non-conservative part of the problem, have also been proposed~\citep{lorenzi2022invasion}. 

\section{Challenges and perspectives} \label{sec:discussion}
PS-PDEs form a powerful modelling tool to account for the complex phenotypic heterogeneity inherent to a population. With their greater adoption in mathematical models of collective cell migration, various challenges have emerged with respect to modelling extensions, analytical results, numerical methods, and the capacity to connect models with experimental data. We conclude this review with a few perspectives.

\bigskip

\begin{trivlist}
\item {\textbf{I. The challenge of cell signalling}}. \psipde models of cell population dynamics often invoke a somewhat hazy definition of phenotype: a generic variable that defines cell behaviour in a phenomenological manner. In practice, phenotype is linked to the internal signalling state and modulated by environmental factors (e.g. extracellular ligands). ABMs can account for this complexity -- e.g. equipping each agent with a detailed ODE model of signalling with the output altering the phenotypic state -- and various open source modelling toolkits have been developed with such capabilities -- see \citep{metzcar2019review} for an overview. Can similar detail be absorbed within continuous models? As described, environmental heterogeneity can be accounted for in a PS-PDE model, by variables for the external factors that regulate growth, movement, and phenotype changes (e.g. \citealt{ayati2006computational,dyson2007age,fiandaca2021mathematical,lorenzi2018role,villa2021modeling,fiandaca2022phenotype,lorenzi2022trade,lorenzi2023derivation}). However, these usually take a black box approach for the intracellular processes that link the extracellular to cell behaviour. More direct definitions for the variability offer a more explicit connection, e.g. associating heterogeneity with tumbling bias in PS-PDE models for {\it E.coli} migration \citep{mattingly2022collective,phan2024direct}. Fundamental approaches to link structure to signalling could exploit ABM to population-level formulations, e.g. the framework described in Section~\ref{sec:tools:derivation}; a step here has been taken by \cite{freingruber2024inprogress}, who derived a PS-PDE model wherein phenotype is associated to bound receptor levels, which in turn evolve through the modelling of receptor-ligand binding. PS-PDE models have been formulated where structuring is according to intracellular signalling pathways connected with EMT\footnote{Epithelial-to-mesenchymal transition, where a cells transitions from an epithelial (e.g. high cell-cell adhesion, low motility) to mesenchymal (e.g. cell-extracellular matrix interactions, high motility) state.} processes \citep{guilberteau2023integrative}, although these do not include spatial movement. Others have addressed a similar challenge with different frameworks, for example by starting from a `mesoscopic' kinetic equation that includes dependency on an internal state variable and employing scaling methods to obtain macroscopic PDE models (but not PS-PDEs) with terms that depend on the internal signalling state (e.g. \citealt{erban2004individual,engwer2015glioma}). Nevertheless, this remains an area for significant work, and will become increasingly necessary if this biological detail is required. 
\item[]
\item \textbf{II. Accounting for energy budgets}. Behaviours – migrating, proliferating, protein synthesis, etc -- have energy costs, leading to upper bounds on cell functioning according to the rates at which oxygen, nutrients, etc are absorbed and converted into energy. The links between the energy state and phenotype have received significant attention in recent years. For example, in {\em E. coli} bacteria, much recent attention has explored the trade-offs due to energetic costs \citep{keegstra2022ecological}, such as negative correlations between chemotactic gene promoters and population growth rate \citep{ni2020growth}. EMT processes, associated with stages of tissue development, wound healing, and cancer progression \citep{nieto2016emt,vilchez2021decoding}, have also been scrutinised from an energetic perspective -- e.g. intermediate states that may correspond to local minima within an energetic landscape \citep{tam2013epigenetics,zadran2014surprisal}. Energy, in a broad sense, can be accounted for through modelling the environmental sources (oxygen, nutrients, etc), as described in {\bf {I.}} To describe the subsequent conversion of these sources by the cell returns us to the above discussion on intracellular signalling.  
\item[]
\item {\textbf{III. Extension of other biological movement models}}. In Section \ref{pdetopspide} we have illustrated the extension of a number of common PDE models of biological movement into the PS-PDE form, specifically diffusion-based, pressure-based, and taxis-based models described therein. There are, naturally, further models used to describe different forms of biological movement. Non-local aggregation models \citep{painter2024biological}, wherein advection through physical space is according to a non-local evaluation of surrounding population density, have become a popular tool to describe the impact of direct interactions -- such as cell-cell adhesion -- on movement. Variation in adhesion is a common and potent form of phenotypic variation: downregulated cell-cell adhesion can lead to more invasive phenotypes within cancer invasion \citep{vilchez2021decoding}; differential adhesion is a powerful driver of cellular sorting during tissue patterning \citep{tsai2022adhesion}. While non-local aggregation models have considered cellular heterogeneity in a discrete manner (e.g. two populations with distinct adhesion, see \citep{painter2024biological} and references therein), extensions to continuous heterogeneity are natural; steps in this direction have been made, for example see \citep{engwer2017structured,lorenzi2024phenotype}. The resulting equations, though, present a formidable mathematical and numerical challenge: for example, a doubly non-local structure, with integral terms for summation across both space and phenotype.
\item[]
\item {\bf IV. Discrete, continuous, or both?} As noted several times, a natural application of \psipdes would be to describe follower-leader type behaviour, for example as can be found in invading cancer cells \citep{vilchez2021decoding}, neural crest migration \citep{mclennan2015neural}, and wound healing \citep{vishwakarma2020dynamic}. While a division into followers and leaders is conceptually straightforward, this can be overly simplistic and misleading \citep{theveneau2017leaders}. A broad spectrum of states may exist between `follower' and `leader', with transitions in between. \psipde models, therefore, can form a natural framework to describe this complexity. But questions then arise regarding the appropriateness of using continuous densities: within such systems, certain states may be confined to small subpopulations -- for example, an invasion process driven by relatively few cells at the front. Turning to a fully discrete representation is one possibility, but this may also be suboptimal if other states are exhibited by vast numbers of cells. How can we form a bridge, spanning discrete to continuous populations according to phenotypic state? Some hybrid approaches offer tantalising directions, e.g. allowing individual cells to emerge from or merge with some continuous distribution, in a way that ensures mass is conserved \citep{chiari2022hybrid}.
\item[]
\item {\bf V. Self-organisation within phenotype-structured populations.} \psipdes are sophisticated equations and present a formidable mathematical challenge. However, in-roads are being made: for example, we illustrated the growing arsenal of techniques to understand phenotype-structured travelling waves. Travelling waves are one well-known dynamic that can arise from PDEs, another being self-organising phenomena. \cite{turing1952} laid the essential groundwork here, via his well known model for morphogenesis. The now standard stability analysis relied on a near homogeneous initial state but, as noted earlier, populations are rarely homogeneous and one could expect phenotypic heterogeneity across even a superficially similar population -- e.g. varying levels of signal activity. Extending pattern formation analyses to models that contain phenotype structuring is relatively straightforward under simpler binary phenotypes -- e.g. see \citep{painter2009continuous,pham2012density,macfarlane2022impact} -- as the resultant PDEs are of low order and the same standard Turing-type analysis can be applied. But extending these analyses to the non-local PS-PDE framework has, to today, received little attention.   
\item[]
\item {\bf VI. Analysis of qualitative and quantitative properties of solutions.} The analysis of PS-PDEs modelling spatial spread and evolutionary dynamics of forms~\eqref{structuredFKPP}, \eqref{structuredPM}, and~\eqref{structuredKS} is a relatively young research field, wherein a number of fascinating and challenging open problems have emerged. Just to mention a few problems related to aspects we have focused on in this review, neither explicit nor semi-explicit solutions are known, with the exception of the case discussed in Section~\ref{sec:analysis:TWCPPSPIDE}. This limits our current level of information on the quantitative properties of the solutions to these models. Moreover, formal asymptotic results on concentration phenomena across travelling waves, like those discussed here, have still to be established rigorously. In this we anticipate major difficulties to be posed due to the lack of compactness~\citep{mirrahimi2015asymptotic,jabin2023selection}. Furthermore, there are no rigorous asymptotic results, of the type presented by~\cite{david2023phenotypic}, on the derivation of free-boundary problems of Hele-Shaw type from PS-PDE models of form~\eqref{structuredPM}, when the mobility parameter is a function of the phenotypic state. This problem appears to be far from being closed given that, as of today, related problems remain open even for systems of PDEs with cross-diffusion terms corresponding to the case where the phenotypic state is binary~\citep{lorenzi2016interfaces,david2024degenerate}. Solving these and related problems will entail harnessing a  range of tools and techniques from across different research areas for the analysis of non-linear and non-local PDEs, thus promoting cross-fertilisation of these areas. Further, it may lead to the development of new mathematical methods that could also be transferable to cognate research fields.
\item[]
\item 
{\bf VII. \psipdes over networks. } Migration not only allows cells to infiltrate neighbouring areas but also disperse to distant sites, as happens in the metastatic spread of cancer. The question of whether cancerous cells that originate in one organ can migrate to, adapt to, and persist within another organ is of manifest interest; related questions arise in an ecological context. In such instances, continuous descriptions of space may no longer be appropriate -- the distance between two organs is not easily defined -- and formulations on a network may be more appropriate. Steps in this direction have been made, by describing the dynamics of the phenotype density in each network node with a non-local PDE of  form~\eqref{nonlocalFKPP}, and adding sink/source terms for the migration across adjacent edges. Assuming fitness functions in the form of~\eqref{def:REx1},\eqref{def:pEx1}, analytical results extend those outlined in Section~\ref{sec:analysis:CPNPDE} and highlight that spatial movement — even in discrete space settings — may provide the substrate for polymorphism, both under constant~\citep{mirrahimi2013adaptation,mirrahimi2020evolution} or phenotype-dependent migration rates~\citep{padovano2024development}. 
The long-term persistence (or extinction) of phenotype-structured populations in such modelling framework was investigated by~\cite{hamel2021adaptation,alfaro2023adaptation}, in the context of host-pathogen interactions.
Nevertheless, \psipdes over networks currently neglect explicit spatial dynamics within each node -- with the exception of a graph-like description of the habitat in each node~\citep{boussange2022eco} -- and the inclusion of continuous spatial structures for node-specific dynamics is an open research avenue.

\item[]
\item {\bf VIII. Development and analysis of numerical schemes.} 
Relying on structuring over both phenotype and physical space, \psipde models necessitate discretisation over multiple dimensions to be simulated. This aspect, coupled with the presence of non-local and highly non-linear terms, often makes solving numerically the equations comprised in these models prohibitively computationally expensive, especially when high-dimensional phenotypic and physical domains are simultaneously considered. As such, a significant challenge is posed when it comes to developing efficient algorithms implementing accurate numerical schemes. As discussed in Section~\ref{sec:tools:numerics}, approaches to date have typically invoked semi-classical ideas (method of lines, finite difference, finite volume schemes, etc), which benefit from a substantial literature and flexibility. However, whether other approaches -- such as pseudospectral methods \citep{fornberg1998practical}, discontinuous Galerkin methods \citep{cockburn2012discontinuous}, particle methods~\citep{alvarez2023particle}, and deep-learning algorithms~\citep{boussange2023deep} -- could be used to generate more efficient algorithms is an intriguing possibility. Beyond the development of efficient algorithms, systematic numerical analysis of discretisation schemes for \psipde models of cell movement of the type considered here is generally lacking in the extant literature, which would provide valuable information on accuracy of numerical solutions; this is certainly another avenue for future research. 
\item[]
\item {\bf IX. Bridging the gap between models and data.} 
Recent technological advances have made it possible to access a vast amount of data on the phenotypic properties of cells, both at the single-cell level and across space (see Figure~\ref{schematicfigure}(c)), that could be used to validate \psipde models of collective cell migration. While some works already leveraged proteomics data from bulk measurements, of the type shown in panel (ii) of Figure~\ref{schematicfigure}(c), for parameter estimation in spatially homogeneous models of phenotypically-structured cell population dynamics~\citep{almeida2024evolutionary,celora2023spatio}, this challenge is yet to be tackled in \psipdes models of cell movement. Indeed spatial data come at varying levels of resolution and may be more prone to stochastic variation compared to bulk measurement data, making the task of PDE calibration more involved. This raises the question of practical identifiability of parameters in \psipde models, which is bound to merge complexities found when addressing this in spatially-structured phenotypically homogeneous  populations~\citep{liu2024parameter} and in well-mixed phenotype-structured populations~\citep{browning2024identifiability}.

\end{trivlist}

\bmhead{Acknowledgements} The authors would like to thank F.R. Macfarlane for her help with Figure~\ref{pressurebasedfigure}. TL and KJP are members of INdAM-GNFM. KJP acknowledges “Miur-Dipartimento di Eccellenza” funding to the Dipartimento di Scienze, Progetto e Politiche del Territorio (DIST). CV is a Fellow of the Paris Region Fellowship Programme, supported by the Paris Region. 

\bmhead{Funding} 
TL gratefully acknowledges support from the Italian Ministry of University and Research (MUR) through the grant PRIN 2020 project (No. 2020JLWP23) ``Integrated Mathematical Approaches to Socio-Epidemiological Dynamics'' (CUP: E15F21005420006) and the grant PRIN2022-PNRR project (No. P2022Z7ZAJ) ``A Unitary Mathematical Framework for Modelling Muscular Dystrophies'' (CUP: E53D23018070001) funded by the European Union – NextGenerationEU. TL and CV acknowledge also support from the CNRS International Research Project ``Modelisation de la biomecanique cellulaire et tissulaire'' (MOCETIBI). This project has received funding from the European Union’s Horizon 2020 research and innovation programme under the Marie Skłodowska-Curie grant agreement No 945298-ParisRegionFP. 

\bmhead{Competing interests} The authors have no competing interests to declare that are relevant to the content of this article.

\bmhead{Ethical Statement} Not applicable.

\bmhead{Consent for publication} All authors have given approval for publication.

\bmhead{Data availability} Not applicable.

\bmhead{Materials availability} Not applicable.

\bmhead{Code availability} Python code is available on the GitHub repository \url{https://github.com/ChiaraVilla/LorenziEtAl2025Phenotype} under GNU General Public License (\url{https://www.gnu.org/licenses/}).

\bmhead{Author contributions} All authors contributed equally to this work.



\bibliography{references}

\begin{thebibliography}{254}
\bibitem[{Malthus(1798)}]{malthus1798}
Malthus TR (1798) An essay on the principle of population, as it affects the
  future improvement of society. {W}ith remarks on the speculations of {M}r.
  {G}odwin, {M}. {C}ondorcet and other Writers. J. Johnson, London
\end{thebibliography}

\end{document}